\documentclass[aps,prd,twocolumn,nofootinbib]{revtex4}
\usepackage{graphicx}
\usepackage{hhline}
\usepackage[mathscr]{euscript}
\usepackage{outlines}
\usepackage{xcolor}
\usepackage{placeins}
\usepackage{amssymb, amsmath}

\newcommand{\Ms}{M_{\odot}}

\newcommand{\ROP}{R\"{O}P}
\usepackage{tikz}
\usepackage{pgfplots}
\usepackage{hyperref}

\newcommand{\mnras}{Mon.\ Not.\ Roy.\ Astron.\ Soc.}

\newcommand{\apjl}{Astrophys.\ J.\ Lett.}
\newcommand{\aap}{Astron.\ and Astrophys.}

\newcommand{\araa}{Annual Reviews of Astr.\ and Astrophys.}

\usepackage[normalem]{ulem}

\begin{document}

\title{Realistic Finite-Temperature Effects in Neutron Star Merger Simulations}

\author{Carolyn A. Raithel,$^{1,2,3}$ Vasileios Paschalidis,$^{4,5}$ Feryal \"Ozel$^4$}
\affiliation{$^1$School of Natural Sciences, Institute for Advanced Study, 1 Einstein Drive, Princeton, NJ 08540, USA}
\affiliation{$^2$Princeton Center for Theoretical Science, Jadwin Hall, Princeton University, Princeton, NJ 08540, USA}
\affiliation{$^3$Princeton Gravity Initiative, Jadwin Hall, Princeton University, Princeton, NJ 08540, USA}
\affiliation{$^4$Department of Astronomy and Steward Observatory, University of Arizona, 933 N. Cherry Avenue, Tucson, Arizona 85721, USA}
\affiliation{$^5$Department of Physics, University of Arizona, 1118 E. Fourth Street, Arizona 85721, USA}

\begin{abstract}
Binary neutron star mergers provide a unique probe of the dense-matter
equation of state (EoS) across a wide range of parameter space, from
the zero-temperature EoS during the inspiral to the high-temperature
EoS following the merger. In this paper, we implement a new model for
calculating parametrized finite-temperature EoS effects into numerical
relativity simulations. This ``$M^*$-model" is based on a
two-parameter approximation of the particle effective mass and
includes the leading-order effects of degeneracy in the thermal
pressure and energy.  We test our numerical implementation by
performing evolutions of rotating single stars with zero- and non-zero
temperature gradients, as well as evolutions of binary neutron star
mergers. We find that our new finite-temperature EoS implementation
can support stable stars over many dynamical timescales. We also
perform a first parameter study to explore the role of the $M^*$
parameters in binary neutron star merger simulations. All simulations
start from identical initial data with identical cold EoSs, and differ
only in the thermal part of the EoS. We find that both the thermal
profile of the remnant and the post-merger gravitational wave signal
depend on the choice of $M^*$ parameters, but that the total merger
ejecta depends only weakly on the finite-temperature
part of the EoS, across a wide range of parameters.
Our simulations provide a first step toward understanding
how the finite-temperature properties of dense matter may affect
future observations of binary neutron star mergers.
\end{abstract}

\maketitle

\section{Introduction}

With the recent detections of gravitational waves from multiple likely
neutron star-neutron star (NSNS) mergers
\citep{Abbott2017a,Abbott2020}, we are now in a new era of
gravitational wave and multimessenger astronomy. Already, these
gravitational waves have been used to constrain the
properties of the dense nuclear matter contained in the neutron star interior
 \citep[see, e.g.,][for recent
   reviews]{Baiotti2019,Raithel2019a,Chatziioannou2020}. Because the
LIGO-Virgo sensitivity is highest for frequencies $\lesssim$1~kHz, the
main gravitational wave information detected so far comes from the
binary inspiral, during which neutron stars are expected to remain
thermodynamically cold. As a result, all analyses of LIGO-Virgo events
to date have specifically constrained the equation of state (EoS) at
effectively zero temperature.

Following the merger, shock heating is expected to raise the
temperature of the system to 10-100~MeV \citep[e.g.,][for
  reviews]{Baiotti2017,Paschalidis2017}, which is well above the Fermi
energy of the matter. At such temperatures, the thermal pressure can
make up a significant fraction of the total pressure and can thus
influence structure of the merger remnant. This, in turn, has been
shown to affect a wide range of NSNS merger properties, from the
lifetime of the merger remnant to the post-merger gravitational wave
(GW) spectrum and the amounts of matter ejected
\citep[e.g.,][]{Oechslin2007,Baiotti2008,Bauswein2010,Bauswein2010a,Sekiguchi2011,Paschalidis2012}. As
a result, observation of these post-merger properties could provide a
new probe into the EoS at {\it finite temperature}.

While a large number of cold, neutron star EoSs have been calculated
in the zero-temperature limit \citep[for reviews,
  see][]{Lattimer2001,Ozel2016}, there exists a much smaller number of
EoSs that are self-consistently calculated at non-zero
temperatures. These ``finite-temperature" EoSs include the well-known
LS model, which is based on a compressible liquid drop model of nuclei
\citep{Lattimer1991}, and the STOS model, which was calculated using
relativistic mean field theory with a Thomas-Fermi approximation
\citep{Shen1998a}. Another $\sim$10 models have been calculated using
a statistical model developed by \citet{Hempel2010} for different
relativistic mean field models and nuclear mass tables, spanning a
wider range of neutron star properties. Additionally, the
\texttt{CompOSE} online directory for neutron star EoS tables has
provided a pathway for groups to easily publish finite-temperature EoS
tables as they become available, and has further increased the number
of available models~\citep{Typel2015} (for a recent review of
finite-temperature EoSs, see \citep{Oertel2017}).

Despite these efforts, the total number of publicly-available
finite-temperature EoS models remains relatively small and they do not
span the full range of possible dense-matter physics. In addition,
some models are not consistent with modern astrophysical
constraints. For example, several of the finite-temperature EoS tables
predict cold neutron star radii of $\gtrsim$13~km (e.g., the NL3, TM1,
DD2, and TMA EoSs; see, e.g., Table 1 of \cite{Fischer2014} and
references therein), which are in tension with the latest constraints
inferred from LMXB observations and from GW170817
\citep{Ozel2016,Baiotti2019,Raithel2019a}.  Critically, there is
currently no implementation of a framework where one can attach a {\it
  realistic} thermal model to {\it any} underlying cold nuclear EoS,
since all existing finite-temperature EoS tables have already assumed
a particular cold component. Having an analytic, parametric framework 
for the thermal physics would be necessary if we hope to one day
infer the properties of finite-temperature matter
from neutron star merger observations.
 Finally, compared to an
analytic framework, these tabulated EoSs add an extra computational
expense to already-expensive numerical simulations.

In order to span a larger range of underlying physics at a low
computational cost, many authors
have turned, instead, to an ad-hoc and analytic approach, in which the energy
density, $\epsilon$, and pressure, $P$, are decomposed according to
\begin{subequations}
\begin{align}
\begin{split}
\label{eq:epstot}
\epsilon & = \epsilon_{\rm cold}  + \epsilon_{\rm th} 
\end{split}\\
\begin{split}
\label{eq:Ptot}
P & = P_{\rm cold} + P_{\rm th},
\end{split}
\end{align}
\end{subequations}
where the subscript ``cold" indicates that the thermodynamic quantity
is calculated at zero-temperature, while the subscript ``th" indicates
the thermal contribution to that quantity. The cold component can be a
microphysical EoS or an agnostic parameterization, and is typically
assumed to be in $\beta$-equilibrium. A thermal correction is then
added to the cold component, in order to allow for shock heating in
the system. In the so-called ``hybrid" approach, which was first
introduced in~\citep{Janka1993} and is now widely used, the thermal
correction is approximated as
\begin{equation}
\label{eq:hybrid}
P_{\rm th} = \epsilon_{\rm th} \left( \Gamma_{\rm th} - 1 \right),
\end{equation}
where the thermal index, $\Gamma_{\rm th}$, is assumed to be constant with a value that is independent of the cold EoS. 

In certain regimes, such as for an ideal fluid or for a gas of
relativistic particles, the thermal index is indeed constant. In fact,
the values of $\Gamma_{\rm th}$ that are commonly used in recent
binary neutron star simulations, $\Gamma_{\rm th}\in[1.5, 2]$, are
approximately consistent with an ideal-fluid EoS, for which
$\Gamma_{\rm th}=5/3$. This is why the hybrid approach is sometimes
referred to as an ideal-fluid approximation. However, for the
degenerate matter that is expected to be found in the cores of neutron
stars, $\Gamma_{\rm th}$ has a strong density dependence, which is
neglected in this hybrid approach \citep[see,
  e.g.,][]{Constantinou2015a}. By neglecting the effects of
degeneracy, the hybrid approach has been shown to overestimate the
thermal pressure by up to four orders of magnitude at densities of
interest \citep{Raithel2019}, and can introduce significant shifts
into the post-merger gravitational wave frequencies found in NSNS
simulations \citep[][]{Bauswein2010,Figura2020}.

Within Landau's Fermi liquid theory, the density-dependence of
$\Gamma_{\rm th}$ can be written directly in terms of the particle
effective mass \citep{Baym1991,Constantinou2015}. Using this fact,
the authors in \citep{Raithel2019} (hereafter \ROP) introduced a framework for
calculating finite-temperature effects based on a new parametrization
of the particle effective mass, which is referred to as the
$M^*$-approximation. This two-parameter model allows for a robust
calculation of the thermal pressure including the leading-order effects of
degeneracy, while still keeping the flexibility of
Eqs.~\eqref{eq:epstot}-\eqref{eq:Ptot}. As with the hybrid approach,
the $M^*$-approximation of the thermal pressure can be added to any
cold EoS, whether it is microphysical or parametric in nature. This
framework for calculating the EoS at arbitrary temperatures and proton
fractions was found to closely approximate the results of a large
family of EoS tables, with errors of $\lesssim30\%$ in the thermal
pressure at densities of interest (cf. the four orders-of-magnitude
errors of the hybrid approach) \citep{Raithel2019}.

While other frameworks for calculating
the EoS in terms of the particle effective mass
have been proposed 
\citep[e.g.,][]{Schneider2017,Carbone2019,Huth2020,Keller2020},
these models depend on a much larger number of parameters, which
dramatically increases the computational cost of exploring
their parameter spaces with NSNS merger simulations. 
By capturing the relevant thermal physics with just two free parameters,
the $M^*$-approximation makes it
computationally possible to study the role of each parameter in merger
simulations in full numerical relativity. Additionally, because 
the $M^*$-approximation can be combined with any cold EoS,
it becomes possible to explore any part of the full EoS parameter space 
within this framework.

In this paper, we implement the $M^*$-framework for calculating
finite-temperature effects into neutron star merger simulations in
full general relativity. We test the implementation and performance of
the $M^*$-framework in evolutions of isolated rotating stars in
equilibrium, with both zero and non-zero initial temperature profiles,
as well as in full evolutions of NSNS mergers. In all cases, we find
that our implementation of the $M^*$-framework maintains the stable
equilibrium of stars over many dynamical timescales. We also perform a
parameter study to explore the range of outcomes from select NSNS
mergers with different values of $M^*$-parameters. In particular, we
study four sets of $M^*$-parameters which span a broad range of
possible nuclear physics, and we compare the evolutions with these
$M^*$-parameters to evolutions with constant values of $\Gamma_{\rm
  th}$, to demonstrate the differences between the $M^*$- and hybrid
approaches. We find that the inspiral phase and the time to merger are
unaffected by the choice of $M^*$-parameters, but that the thermal
profile of the remnant and the post-merger GW signal are both
sensitive to finite-temperature effects.  We find no numerically
significant difference in the total amounts of matter ejected for the
various $M^*$-parameters explored in this work, although the ejecta
can be a factor of a few lower for $\Gamma_{\rm th}=2$,
compared to any of the $M^*$ evolutions or the hybrid evolution with
$\Gamma_{\rm th}=1.5$.

The structure of the paper is as follows: We start in
Sec.~\ref{sec:EOSuncertainties} with a brief discussion of the current
uncertainties in the finite-temperature EoS.
Section~\ref{sec:outline} presents an overview of the tests performed
in this paper. In Sec.~\ref{sec:methods}, we discuss the numerical
methods used in our simulations, with the implementation of the
$M^*$-framework discussed in detail in Sec.~\ref{sec:Mstar}. Finally,
in Sec.~\ref{sec:results}, we present the results from the NSNS merger
simulations, and we discuss how different assumptions about the
thermal physics affect various merger properties. Convergence tests
and resolution studies can be found in
Appendices~\ref{sec:singlestars} and \ref{sec:BNSresolution}. Unless
otherwise specified, we adopt geometrized units in which $G=c=1$.

\section{Uncertainties in the finite-temperature EoS for dense nuclear matter}
\label{sec:EOSuncertainties}
  
\begin{figure}[!ht]
 \centering
 \includegraphics[width=0.45\textwidth]{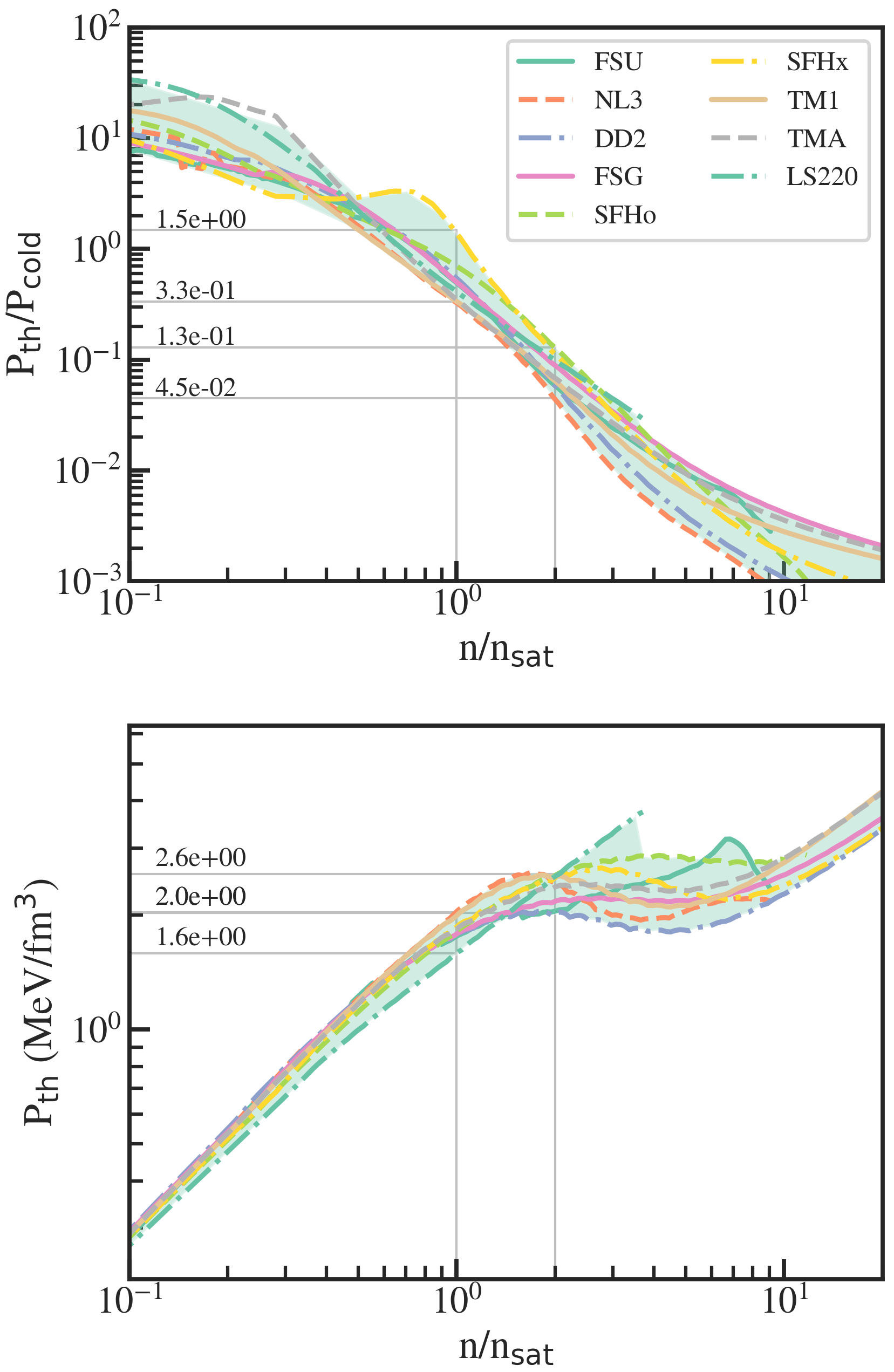}
 \caption{Top: Ratio of thermal-to-cold pressure as a function of the
   density for various finite-temperature EoSs. Bottom: magnitude of
   the thermal pressure for the same EoSs. For each EoS, the thermal
   pressure is computed at $k_B T=20$ MeV and $P_{\rm
     cold}$ is at $k_B T=0.1$ MeV, for proton fraction of $Y_p=0.1$.
     The vertical lines correspond to the nuclear
   saturation density, $n_{\rm sat} = 0.16~\rm{fm}^{-3}$, and $2n_{\rm
     sat}$, while the horizontal lines indicate the maximum range
     in the tabulated EoSs at these densities. The green shading
     is included to visually highlight the range in thermal
     pressures spanned by these EoSs.
   \label{fig:PthoPcold}  }
\end{figure}

Existing finite-temperature EoS tables remain quite uncertain at the
supranuclear densities and high temperatures relevant to binary
neutron star mergers.  The range of thermal pressures predicted by a
sample of commonly used finite-temperature EoSs is shown in
Fig.~\ref{fig:PthoPcold}. These EoSs include the DD2, TMA, TM1, FSG
models calculated within the statistical framework of
\citet{Hempel2010} (and references therein), SFHo and SFHx
\citep{Steiner2013}, NL3 and FSU \citep{Shen2011a}, and the LS220 model of
\citet{Lattimer1991}.  The top panel shows the thermal pressure
relative to the cold pressure, for matter at a temperature of
$k_B T=20$~MeV and proton fraction $Y_e=0.1$.  We note that the ``cold"
pressure corresponds to $k_B T=0.1$~MeV, which is among the lowest
realiable temperatures from the tabulated EoSs, and is
thermodynamically cold in that the temperature is much less than the
Fermi energy of nucleons. The bottom panel of Fig.~\ref{fig:PthoPcold}
shows the magnitude of the thermal pressure under the same
conditions. For these EoSs, the thermal pressure can significantly
exceed the cold pressure at low densities; $P_{\rm th}$ can be
comparable to $P_{\rm cold}$ at the nuclear saturation density
($n_{\rm sat}=0.16$~fm$^{-3}$); and $P_{\rm th}$ is still $\sim10$\%
of the cold pressure at $2n_{\rm sat}$.  Moreover, there is
significant variation between the tabulated EoSs, with the ratio of
$P_{\rm th}/P_{\rm cold}$ differing by a factor of 5 at $n_{\rm sat}$
and by a factor of 3 at $2n_{\rm sat}$, between these EoSs.  We also
note that the temperature after a binary neutron star merger can reach
even higher values than those considered here, with temperatures up to
40-50~MeV at 2-3$n_{\rm sat}$, in which case the thermal pressure can
be up to $\sim 50\%$ of the cold pressure at supranuclear densities,
as we show in Sec.~\ref{sec:thermal}.

Thus, even within the family of commonly-used EoS tables, thermal
effects remain quite uncertain. This uncertainty may be reduced
through observations of neutron stars at high temperatures, such as
during the late stages of a binary neutron star merger. However,
constraining the finite-temperature part of the EoS requires one to be
able to untangle the role of the cold EoS, which is uncertain in its
own right, from any thermal effects. This is not straightforward in
simulations adopting tabulated finite-temperature EoSs,
e.g.,~\cite{Sekiguchi2011,Palenzuela:2015dqa,Radice:2016dwd,Lehner2016a,Foucart:2016rxm,Bovard:2017mvn,Most:2018eaw}. However,
the analytic framework of the $M^*$-approximation, with its
physically-motivated parameters that can be varied systematically and
independently of the cold EoS, provides one major step forward toward
the goal of constraining the finite-temperature EoS with future
observations of NSNS mergers.  We begin to explore this approach in
this work.

\section{Overview of simulations performed}
\label{sec:outline}
In this paper, we implement the $M^*$-approximation into NSNS merger
simulations in full numerical relativity. In order to validate the
implementation and performance of the $M^*$-EoS, we run three
different types of tests. For each test, we evolve the initial data
with: 
\begin{enumerate}
\item the hybrid approximation with a constant $\Gamma_{\rm th}$, and
\item the $M^*$-approximation,
\end{enumerate} 
each added to the same cold EoS.

\begin{table*}
\centering
\begin{tabular}{lccclll}
\hline \hline
Configuration & Gravitational mass & Initial temperature &  Cold EoS && Thermal treatment  \\
\hline 
\vspace{-0.1cm}  			 & &    			    && ~~~~~&$\Gamma_{\rm th}=1.66$& \\  \vspace{-0.3cm}  	
 			Single star & 1.4~$\Ms$  &	$P_{\rm th}=0$  & $\Gamma=2$ polytrope&  &  \\
 						&   &   			    &&& $M^* (n_0=0.12~\text{fm}^{-3}, \alpha=0.8)$  \\ \hline

\vspace{-0.1cm}		 	 &  &    	 		 &&& $\Gamma_{\rm th}=1.66$ \\ \vspace{-0.3cm}  	
 			Single star & 1.4~$\Ms$  &	$P_{\rm th}=0.1 P_{\rm cold}$ & $\Gamma=2$ polytrope&   &  \\
 						 && 			    	  &&& $M^* (n_0=0.12~\text{fm}^{-3}, \alpha=0.8)$ \\ \hline

  \vspace{-0.05cm}			& &     			  &&& $\Gamma_{\rm th}=1.5$ \\
 \vspace{-0.05cm}			& &    	 		 &&& $\Gamma_{\rm th}=2$  \\ \vspace{-0.35cm}  	
 				NSNS	& 1.4~$\Ms+1.4~\Ms$ & 	$P_{\rm th}=0$ & ENG (piecewise polytropes) & &&  \\
 						& &    			  &&& $M^* (n_0=0.08~\text{fm}^{-3}, \alpha=0.6)$  \\
 						& &    			  &&& $M^* (n_0=0.08~\text{fm}^{-3}, \alpha=1.3)$  \\
 						& &   			  &&& $M^* (n_0=0.22~\text{fm}^{-3}, \alpha=0.6)$ \\
						& &   			  &&& $M^* (n_0=0.22~\text{fm}^{-3}, \alpha=1.3)$ \\
\hline
\end{tabular}
\caption{Summary of tests run. The parameters $n_0$ and $\alpha$ in the $M^*$ model are described below.}
  \label{table:simulations}
\end{table*}

In the first set of tests, we evolve a single rotating, cold neutron
star, in order to ensure that the star remains cold over time. In the second,
we evolve a rotating, single neutron star, to which we add a non-zero
temperature gradient. By studying whether the temperature gradient can
be maintained without loss of stability and without spurious growth,
this provides a more stringent test of the $M^*$-EoS. Finally, we
evolve a set of NSNS mergers with a large range of
$M^*$-parameters. This enables us to study the performance of the
$M^*$-EoS in a dynamical setting, in which the stars start cold and
develop significant temperature gradients through
shock-heating. Additionally, by using a wide range of
$M^*$-parameters, we perform an initial parameter study of
how each $M^*$-parameter affects the late-stage properties of an NSNS
merger and we compare the outcomes to the 
ideal-fluid approximation. We summarize the various tests run in
Table~\ref{table:simulations}.

\section{Numerical Methods}
\label{sec:methods}

All simulations were performed with the Illinois dynamical spacetime,
general-relativistic magnetohydrodynamics (GRMHD),
adaptive-mesh-refinement code, which has most recently been described
in \citet{Etienne2015}, and is built within the Cactus/Carpet
framework \citep{Allen2001,Schnetter2004,Schnetter2006}. The spacetime
is evolved using the Baumgarte-Shapiro-Shibata-Nakamura formulation of
the Einstein equations \citep{Shibata1995,Baumgarte1999}. We use 1+log
time slicing of the lapse \citep{Bona1995} and a 2nd-order
``Gamma-driver" condition for the shift \citep{Alcubierre2003}.
Additionally, we modified the original Illinois GRMHD
code to use the primitive variable recovery routine described in
\citet{East2012}.
  
\subsection{The $M^*$-approximation of thermal effects}
\label{sec:Mstar}
During the evolutions, the EoS is calculated locally at each time
step. The total energy and pressure are taken to be the sum of a cold
component and a thermal component, as in
Eqs.~\eqref{eq:epstot}-\eqref{eq:Ptot}. For the hybrid evolutions, the
thermal component is trivially calculated according to
Eq.~\eqref{eq:hybrid}, for constant $\Gamma_{\rm th}$. In the
$M^*$-formalism, the thermal pressure and energy are not so simply
related. In this section, we summarize the $M^*$-framework for
calculating $P_{\rm th}$ and $E_{\rm th}$ from one another, as was first
presented in \ROP.

In this framework, the thermal energy per baryon and the thermal pressure are given by
\begin{widetext}
\begin{subequations}

\begin{align}
\begin{split}
\label{eq:Eth}
E_{\rm th}(n, T, Y_p) = \frac{ 4 \sigma f_s  T^4}{c n} +
			 \left\{ \left(\frac{3 k_B T}{2}\right)^{-1} \right.
	+ \left. \left[ a(0.5 n, M^*_{\rm SM}) + a(Y_p n, m_e) Y_p   \right]^{-1} T^{-2}   \vphantom{\frac{i_i}{i_i}} \right\}^{-1}
\end{split}
\end{align}

\begin{align}
\begin{split}
\label{eq:Pth}
P_{\rm th}(n, T, Y_p) = \frac{4 \sigma f_s  T^4}{3 c} + 
			\left\{ \left(n k_B T\right)^{-1} \vphantom{\frac{i_i}{i_i}} \right.
	\left. -\left[ \frac{\partial a(0.5 n, M^*_{\rm SM})}{\partial n} + \frac{\partial  a(Y_p n, m_e)}{\partial n} Y_p    \right]^{-1} n^{-2} T^{-2}  \right\}^{-1}
\end{split}
\end{align}

\end{subequations}
\end{widetext}
where $n$ is the baryon number density, $T$ is the temperature, $Y_p$
is the proton fraction, $\sigma$ is the Stefan-Boltzmann constant, $c$
is the speed of light, $f_s$ is the number of relativistic species,
$a$ is the level-density parameter, $M^*_{\rm SM}$ is the relativistic
Dirac effective mass of symmetric nuclear matter, and $m_e$ is the
electron mass. The adiabatic sound speed can also be calculated within
this framework, as in Appendix~B of \ROP. We note that in the original
framework of \ROP, there was a typo, such that $M^*_{\rm SM}$ was
incorrectly preceded by a factor of 0.5 in the level-density parameter
term. We have corrected this expression in Eqs.~\eqref{eq:Eth} and
\eqref{eq:Pth} and in the remainder of the present work.

Equations~\eqref{eq:Eth} and \eqref{eq:Pth} each consist of three
terms, which characterize the different density regimes that can be
encountered in an NSNS merger. The first term ($\propto T^4$)
describes the energy of a relativistic gas of particles with $f_s$
degrees of freedom. This term dominates at very low densities and thus
will affect the atmosphere and low-density outflows during a
merger. The second term ($\propto T$) is the ideal fluid contribution,
which dominates at intermediate densities, up to $\sim n_{\rm
  sat}=0.16~\rm{fm}^{-3}$. At higher densities ($\gtrsim n_{\rm sat}$;
although the exact transition density depends sensitively on the
temperature), the matter is degenerate and the corresponding thermal
energy scales as $T^2$ at leading-order. Adding the ideal and
degenerate-limit terms inversely ensures that the degenerate term
dominates at high densities and guarantees a smooth transition between
the ideal and degenerate regimes. We note, however, that doing this
separately in Eqs.~\eqref{eq:Eth} and \eqref{eq:Pth} means that these
quantities are no longer exactly thermodynamically linked across the
narrow range of densities where the transition occurs (for further
discussion, see \ROP).

When calculating the number of relativistic species that contribute to
the thermal energy, we consider two limits. For $k_bT\ll 2 m_ec^2$,
photons are the dominant relativistic species, making $f_s=1$.  For
$k_BT\gg2 m_e c^2$, electrons and positrons become relativistic as
well, each with 7/8 degrees of freedom, and thus $f_s=11/4$. At
temperatures above 10~MeV, thermal neutrinos and anti-neutrinos appear;
 however, following the convention of common
finite-temperature EoS tables, we neglect the thermal contribution
from neutrinos in this calculation, but it is straightforward to 
account for them in our approach. In order to smoothly connect the two
temperature regimes of interest, we approximate the number of
relativistic species with a simple linear interpolation, according to
\begin{equation}
\label{eq:fs}
 f_S = 
\begin{cases}
1,   & k_B T < 0.5 ~ \text{MeV} , \\
-0.75 + 3.5 \left( \frac{k_B T}{1~\rm{MeV}} \right), & 0.5 \le k_B T < 1 ~ \text{MeV}, \\
 11/4,  & k_B T \ge~1 \text{MeV}.
\end{cases}
\end{equation}

At higher densities, the degenerate thermal terms are characterized by
the level-density parameter,
\begin{equation}
\label{eq:a}
a(n_q, M_q^*) \equiv \frac{\pi^2 k_B^2}{2} \frac{ \sqrt{ \left( 3 \pi^2 n_q \right)^{2/3}  (\hbar c)^2 + M_q^{*2} }}{\left( 3 \pi^2 n_q \right)^{2/3} (\hbar c)^2 },
\end{equation}
where $n_q$ and $M_q^*$ are the density and relativistic Dirac effective
mass of the species, respectively. Here, we consider only symmetric
nuclear matter, for which the relevant species are protons, neutrons,
and electrons, and we neglect the small change to the thermal pressure
caused by the matter having unequal numbers of protons and neutrons
(see \ROP~for additional details). In symmetric matter, the number
densities of protons and neutrons are equal by definition (i.e., $n_p
= n_n = 0.5 n$), and we take the neutron and proton
effective masses to be comparable as well, such that $M^*_p \approx
M^*_n \approx M^*_{\rm SM}$, where the last term is the symmetric
matter effective mass. We parametrize the effective mass function as
\begin{equation}
\label{eq:Meff}
M^*_{\rm SM} = \left\{ (m c^2)^{-2} + \left[ mc^2 \left( \frac{n}{n_0 }\right)^{-\alpha} \right]^{-2} \right\}^{-1/2},
\end{equation}
where $n$ is the total baryon number density.

In this parametrization, we fix the low-density baryon mass to the
energy per baryon of $^{56}$Fe, $mc^2=$930.6~MeV. This leaves us with
two free parameters: $n_0$, which controls the density at which
degeneracy becomes significant, and $\alpha$, which controls the rate
at which the effective mass decreases at high densities and which is
related to the strength of the particle interactions in the
matter. These are the parameters that will be varied in our NSNS
evolutions.  The effective mass of the electrons is approximately
constant due to their small interaction cross-section, so their
effective mass simply reduces to the electron mass.

For the bulk of the matter within a neutron star merger remnant, the
neutrino opacity is expected to be large enough that the neutrinos are
trapped on the timescales considered in this paper
\citep{Rosswog2003,Paschalidis2012}.  As a result, the local proton
fraction in the remnant is not changed by neutrino interactions,
although $Y_p$ can deviate from its initial $\beta$-equilibrated value
through advection.  Because $Y_p$ does not enter the hybrid
approximation of Eqs.~\eqref{eq:epstot}-\eqref{eq:Ptot}, it is not
possible to consider the advection of $Y_p$ within that
framework. Thus, in the regime of large neutrino opacities, the hybrid
approximation implicitly requires that the matter remains in its
initial composition (i.e., cold $\beta$-equilibrium), so that the cold
pressure expression does not change. In order to perform the most
direct comparison between the $M^*$- and the hybrid approximation, in
this work we also assume that the matter maintains its initial cold,
$\beta$-equilibrium composition. It should be noted, however, that the
most general form of the $M^*$-formalism allows for full composition
dependence \citep{Raithel2019}.

Accordingly, we set the proton fraction of the matter at each time
step such that it corresponds to that of cold $\beta$-equilibrium. For
nucleonic matter in $\beta$-equilibrium, the proton fraction is
uniquely given by the local density and the symmetry energy, $E_{\rm
  sym}$, according to
\begin{widetext}
\begin{equation}
\label{eq:YpBInv}
Y_{p,\beta}(n) = \frac{1}{2} + \frac{(2 \pi^2)^{1/3}}{32} \frac{n}{\xi} \left\{ (2\pi^2)^{1/3} - \frac{\xi^2}{n} \left[\frac{\hbar c}{E_{\rm sym}(n,T=0)}\right]^3 \right\},
\end{equation}
\end{widetext}
where, for simplicity, we have introduced the auxiliary quantity $\xi$, defined as
\begin{multline}
\label{eq:xi}
\xi \equiv \left[ \frac{E_{\rm sym}(n,T=0)}{\hbar c} \right]^2  \times \\
\left\{ 24 n \left[ 1+ \sqrt{ 1 +  \frac{\pi^2 n}{288}\left(\frac{\hbar c}{E_{\rm sym}(n,T=0) }\right)^3}\right]  \right\}^{1/3}.
\end{multline}  

We parameterize the nuclear symmetry energy in terms of a kinetic and potential-like term \citep[as in][]{Tsang2009, Steiner2010}, according to
\begin{equation}
\label{eq:Esym}
E_{\rm sym}(n, T=0)= \eta E_{\rm sym}^{\rm kin}(n)  
	+ \left[ S_0 - \eta E_{\rm sym}^{\rm kin}(n_{\rm sat}) \right] \left(\frac{n}{n_{\rm sat}}\right)^{\gamma},
\end{equation}
where $S_0$ is the value of the symmetry energy at the nuclear
saturation density. The ``kinetic" term, $E_{\rm sym}^{\rm kin}$
arises from the change in the Fermi energy, $E_F$, of a gas as the
relative densities of protons and neutrons ($n_n$ and $n_p$) change,
and is given by\footnote{We note that there was a factor-of-2 
typo in the equation for $E_{\rm
  sym}^{\rm kin}$ in \ROP~which has been corrected in
Eqs.~\eqref{eq:Ekin} and \eqref{eq:eta}. }
\begin{equation}
\label{eq:Ekin}
E_{\rm sym}^{\rm kin}(n)  = \frac{3}{5} \left[ E_F \left(n_p = n_n =\frac{1}{2} n\right) - E_F(n_n = n)  \right]
\end{equation}
where
\begin{equation}
 E_F(n_q) = \frac{\hbar^2}{2 m} \left( 3\pi^2 n_q \right)^{2/3}.
\end{equation}
 The potential-like term in Eq.~\eqref{eq:Esym}
is less well understood and is, thus, given an arbitrary
density-dependence above the nuclear saturation density, $n_{\rm
  sat}$, through the free parameter $\gamma$. Finally, the parameter
$\eta$, which accounts for the short-range correlations \citep{Xu2011,
  Vidana2011, Lovato2011, Carbone2012, Rios2014,Hen2015}, can be
written as
\begin{equation}
\label{eq:eta}
\eta = \frac{5}{9} \left[ \frac{ L_0-3 S_0 \gamma}{\left(2^{-2/3}-1\right)\left(2/3 - \gamma \right) E_F(n_{\rm sat})} \right],
\end{equation}
where $L_0$ is related to the slope of $E_{\rm sym}$ at $n_{\rm sat}$.

We adopt the full symmetry energy model described 
above for the regime of uniform, nuclear matter, i.e., at 
densities above 0.5$n_{\rm sat}$. At lower densities, however,
this model breaks down. Thus, for $n<0.5 n_{\rm sat}$, we transition
to a function that smoothly decays to zero, such that
the symmetry energy is given by
\begin{multline}
E_{\rm sym}(n) = E_{\rm sym}(0.5 n_{\rm sat}) + \\
 P_{\rm sym}(0.5 n_{\rm sat})\left[ \frac{ \left( \frac{n}{0.5 n_{\rm sat}}\right)^{x-1}-1}{ 0.5 n_{\rm sat} (x-1)} \right], \quad n < 0.5 n_{\rm sat}
\end{multline}
where $x$ is empirically determined to ensure that $Y_{p,\beta}$ rises to 0.5 at low densities,
and where $P_{\rm sym}(n) \equiv n^2 \partial E_{\rm sym}/\partial n$ 
(see \citep{Raithel2019}~for the full expression).
The form of this low-density symmetry energy is chosen to ensure a
 reasonable behavior of $Y_{p, \beta}$ at low densities.

\begin{figure*}[!ht]
\centering
\includegraphics[width=0.95\textwidth]{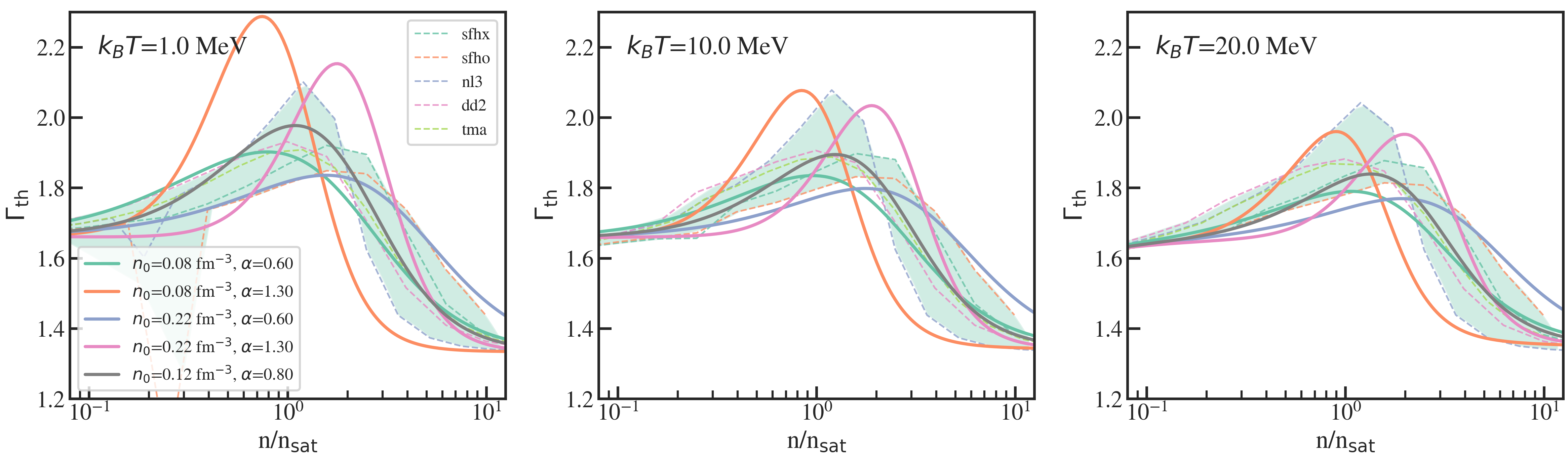}
\caption{\label{fig:params} Effective thermal index for the parameters
  explored in this work at three different temperatures. From left to
  right, the panels show $\Gamma_{\rm th}$ at $k_BT =$1, 10, and
  20~MeV; all panels are calculated for matter in neutrino-less
  $\beta$-equilibrium.  We also show the effective thermal index for a
  sample of finite-temperature EoSs as dotted lines, for comparison.
  The green shading is included to visually represent the range of
  $\Gamma_{\rm th}$ values spanned by the realistic EoS tables.  All
  sets of $M^*$-parameters confirm that $\Gamma_{\rm th}$ indeed
  depends on the density. The degree of density-dependence is directly
  governed by the parameter $\alpha$, while the density at which
  $\Gamma_{\rm th}$ begins to vary is determined by the parameter
  $n_0$.}
\end{figure*}

In this work, we fix the symmetry energy parameters to values that
best fit the SFHo finite-temperature EoS, with $S_0$=31.57~MeV,
$L_0$=47.10~MeV, and $\gamma=0.41$ \citep{Steiner2013a,Raithel2019}.
The SFHo EoS is based on a relativistic mean field theory calculation,
using the statistical model of \citet{Hempel2010}, and is constructed
to be consistent both with experimental nuclear data and astrophysical
observations of neutron stars; additionally, SFHo has similar cold
neutron star properties to ENG, which is the cold EoS used in our
binary neutron star merger calculations (see $\S$\ref{sec:id}).

Finally, we need to be able to convert between the energy and the
total pressure.  Unlike in the hybrid approximation of Eq.~\eqref{eq:hybrid},
Eqs.~\eqref{eq:Eth} and \eqref{eq:Pth} describe multiple regimes, each
of which have a different density- and temperature-dependence. As a
result, there is no simple expression for $P_{\rm th}$ in terms of
$E_{\rm th}$ and vice versa. We can,
nevertheless, simply convert between these quantities as follows:
Given the density, $Y_{p,\beta}(n)$, and one thermodynamic quantity --
either $E_{\rm th}$ or $P_{\rm th}$ -- we use Eq.~\eqref{eq:Eth} or
\eqref{eq:Pth} to numerically solve for the temperature, using a
combination of the Newton-Raphson and bisection methods. We then use
$n$, $Y_{p,\beta}(n)$, and the inverted temperature to directly
calculate the other thermodynamic variable.

When implementing this framework numerically, we also need to adopt
one additional modification. During binary neutron star evolutions,
numerical errors can cause the total pressure to drop below the cold
pressure. By Eq.~(\ref{eq:Ptot}), this would imply the thermal pressure
has become negative; but, negative thermal pressures are not allowed
within the microphysical $M^*$-framework.  To mitigate this unphysical
error, we impose a pressure floor to prevent the thermal pressure from
becoming too negative. It was previously shown in~\citep{Etienne2015}
that setting the pressure floor at exactly $P_{\rm cold}$ can cause
large drifts in the central density of single-star evolutions; as a
result, we adopt an intermediate pressure floor of $0.9P_{\rm
  cold}$. In the regime where the thermal pressure or energy become
negative, we switch to a hybrid EoS with $\Gamma_{\rm th}=2$ in order
to facilitate the conversion between $E_{\rm th}$ and $P_{\rm th}$.
 
Because the $M^*$-framework involves only two free parameters ($n_0$
and $\alpha$; as we are fixing the symmetry energy parameters), we
find that implementing the $M^*$-framework into binary evolution
calculations introduces a slowdown of only $\sim30\%$ to the overall
speed of the code, compared to an identical evolution with the hybrid
approximation.

\subsubsection{$M^*$-parameters explored in this work}

For a sample of nine finite-temperature EoS tables, the
$M^*$-parameters have been found to range between $n_0 \in
[0.10-0.22]~$fm$^{-3}$ and $\alpha \in [0.72-1.08]$, for symmetric
nuclear matter \citep{Raithel2019}. For our single star tests, we use
one representative set of parameters, with $n_0=0.12$~fm$^{-3}$ and
$\alpha=0.8$. For the binary evolutions, we explore values of
$n_0$=0.08 and 0.22~fm$^{-3}$, and $\alpha=0.6$ and 1.3, which
approximately bracket the range found in the sample of tabulated EoSs
cited above. These choices of parameters are summarized in
Table~\ref{table:simulations}.

An effective thermal index for each of these models can be calculated according to
\begin{equation}
\Gamma_{\rm th} = 1+ \left( \frac{P_{\rm th} (n,T, Y_p)}{n E_{\rm th} (n,T, Y_p)} \right).
\end{equation} 
The resulting thermal indices for the five parameter combinations used
in this work are shown in Fig.~\ref{fig:params}. We also include in
Fig.~\ref{fig:params} the thermal index for several finite-temperature
EoS tables, as dotted lines, for comparison. We find a strong density
dependence in the thermal index for all of the $M^*$ EoSs, as
expected. The range of $\Gamma_{\rm th}$ for the four extremal
$M^*$-parameters approximately brackets the range of tabulated
$\Gamma_{\rm th}$, as intended. The set of $M^*$-parameters used for
the single-star test ($n_0=0.12~\text{fm}^{-3}$, $\alpha=0.8$; shown
in gray in Fig.~\ref{fig:params}) was chosen as a more realistic set
of parameters, and it can be seen in Fig.~\ref{fig:params} that this
choice is approximately consistent with the equivalent $\Gamma_{\rm
  th}$ of the tabulated EoSs considered. Figure~\ref{fig:params} also
demonstrates the dependence of $\Gamma_{\rm th}$ on the
$M^*$-parameters: namely, we find that the density at which
$\Gamma_{\rm th}$ starts to vary is directly governed by the parameter
$n_0$, while the degree of density-dependence is determined by the
parameter $\alpha$. Microphysically, we can interpret $n_0$ as being
related to the density at which particle interactions start to become
significant and $\alpha$ as corresponding to the strength of those
particle interactions.

\subsection{Initial conditions}
\label{sec:id}
We now describe the initial conditions for the various tests performed in this paper.
The single-star initial data were constructed using the code of
\citet{Cook1994,Cook1994a}. For both temperature configurations, we
used a $\Gamma=2$ polytrope for the cold EoS and we assumed the matter
was initially in $\beta$-equilibrium, with the proton fraction set
according to Eq.~(\ref{eq:YpBInv}). For the zero-temperature test,
this completely describes the EoS. For the finite-temperature
single-star test, we added a thermal gradient to this cold EoS, such
that the thermal pressure is 10\% of the cold pressure at all
densities. We constructed one EoS table with the $P_{\rm th}/P_{\rm
  cold}=0.1$ profile assuming $\Gamma_{\rm th}=1.66$ to calculate the
associated energies, as well as a second EoS table with the same
thermal pressure profile but instead assuming the $M^*$-approximation
with $n_0=0.12$~fm$^{-3}$ and $\alpha=0.8$. For all single-star tests,
the gravitational mass of the stars was 1.4~$\Ms$ and the stars were
set to be rapidly rotating, such that the ratio of rotational to
gravitational binding energy was $T/W=0.037$, with a ratio of the
polar-to-equatorial radii of 0.85. We note that while the ratio $T/W$ is the same
for the three tests considered here, the individual values of $T$ and $W$ vary
between them. 

The binary neutron star initial data were constructed with the Compact
Object Calculator (\texttt{COCAL}) code
\citep{Uryu2012,Tsokaros2015,Tsokaros2018}. The initial configuration
describes two unmagnetized, equal-mass neutron stars in a
quasi-circular orbit, with an Arnowitt-Deser-Misner (ADM) mass of
2.8~$\Ms$, an initial separation of 35~km, and ADM angular momentum of
$J_{\rm ADM}/{M_{\rm ADM}}^2$= 0.93. The neutron stars start at
zero-temperature and are described by a piecewise polytropic
representation of the ENG EoS (\cite{Engvik1994,Engvik1996}, as fit
for in \cite{Read2009}). With this EoS, the radius of a 1.4~$\Ms$,
non-spinning, cold neutron star is 12.06~km and the corresponding
maximum mass is 2.24~$\Ms$. Both properties are consistent with the
latest astrophysical observations (for a review of neutron star radii,
see e.g., \cite{Ozel2016}; for maximum mass constraints,
\cite{Demorest2010,Antoniadis2013,Fonseca2016,Cromartie2020}). For
the rotating configuration used here, the coordinate equatorial radii
of the initial stars is 13.6~km.

\subsection{Grid hierarchy}
\label{sec:grid}
For the single star evolutions, we use a fixed mesh refinement grid
hierarchy, consisting of 7 refinement levels, each with a 2:1
refinement ratio. The half-side length of the finest level is set to
be 30\% larger than the coordinate equatorial radius of the neutron
star, so that the entire star is contained within the innermost
refinement level. This level has grid spacing such that the equatorial
diameter of the neutron star is covered with 82 points for the
baseline resolution. We also run high-resolution evolutions with half
this grid spacing (i.e., 164 grid points across the star).

For the binary evolutions, we use 9 refinement levels, again each with
a 2:1 refinement ratio. The computational domain extends across
$[-4468,4468]^2 \times [0,4468]$~km. Equatorial symmetry is imposed to
save computational resources. The baseline resolution corresponds to
$\sim100$ points across the diameter of each initial neutron star at
the finest level, with a resolution of $dx_{\rm finest}\approx
0.27$~km. We also perform simulations at 1.5625$\times$ 
and 2$\times$ the baseline resolution for the $M^*$-EoS with 
$n_0=0.08$~fm$^{-3}$ and $\alpha=1.3$ (i.e., using $\sim150$ 
and 200 points across the diameter of each initial star, respectively).

\subsection{Diagnostics}
\label{sec:diag}
We use several diagnostic quantities to analyze the simulation
output. For all evolutions, we monitor the L2 norm of the Hamiltonian
constraint, $||\mathcal{H}||$, in order to validate our numerical
calculations. We also track the evolution of the maximum rest-mass
density in order monitor the stability of the stars against
gravitational collapse.

Additionally, we extract gravitational radiation using the
Newman-Penrose Weyl scalar $\psi_4$, which is related to the GW strain
via $\psi_4 = \ddot{h}_+ - i \ddot{h}_{\times}$. The Weyl scalar is
decomposed on spheres at large radii ($r\ge120~M$) into $s=-2$
spin-weighted spherical harmonics, such that
\begin{equation}
 \psi_4(t,r,\theta,\phi) = \sum_{\ell=2}^{\infty} \sum_{m=-\ell}^{\ell} \psi_4^{\ell m} (t,r) _{-2}Y_{\ell m}(\theta,\phi)
\end{equation}
where $\theta$ and $\phi$ are defined with respect to angular momentum
axis, $r$ is the extraction radius, and $t$ is the time. The total
strain, $h\equiv h_+ - ih_{\times}$, is then given by
\begin{equation}
\label{eq:ht}
h(t,r,\theta,\phi) = \int_{-\infty}^t dt' \int_{-\infty}^{t'} dt'' \psi_4(t'',r,\theta,\phi).
\end{equation}
We calculate the double time integration using the fixed-frequency
integration (FFI) method~\citep{Reisswig2011}.

Finally, we calculate the amount of matter ejected during the NSNS
evolutions by integrating the total rest-mass density, $\rho_b$,
outside of a given radius $r$ and for matter for which $-u_t>1$ ,
according to
\begin{equation}
\label{eq:Mej}
M_{\rm ej} (>r) = \int_{>r} \rho_b u^t \sqrt{-g} d^3 x,
\end{equation}
where $u^{t}$ is the time-component of the fluid 4-velocity and $g$ is
the determinant of the metric.

\section{Simulation results}
\label{sec:results}

We now turn to the results of our numerical simulations. We start with
a brief summary of the findings from the single star evolutions (for
further details, see Appendix~\ref{sec:singlestars}). We find that
rotating stars evolved with the $M^*$-EoS indeed maintain their
initial temperature profile and remain stable, for both cold and
finite-temperature initial data. Additionally, both $M^*$ evolutions
exhibit second-order convergence in the central rest-mass density over
time, as expected from our numerical scheme.  With this validation of our
numerical methods now in hand, we devote the remainder of this section
to the results of the binary star evolutions.

\begin{figure}[ht]
\centering
\includegraphics[width=0.45\textwidth]{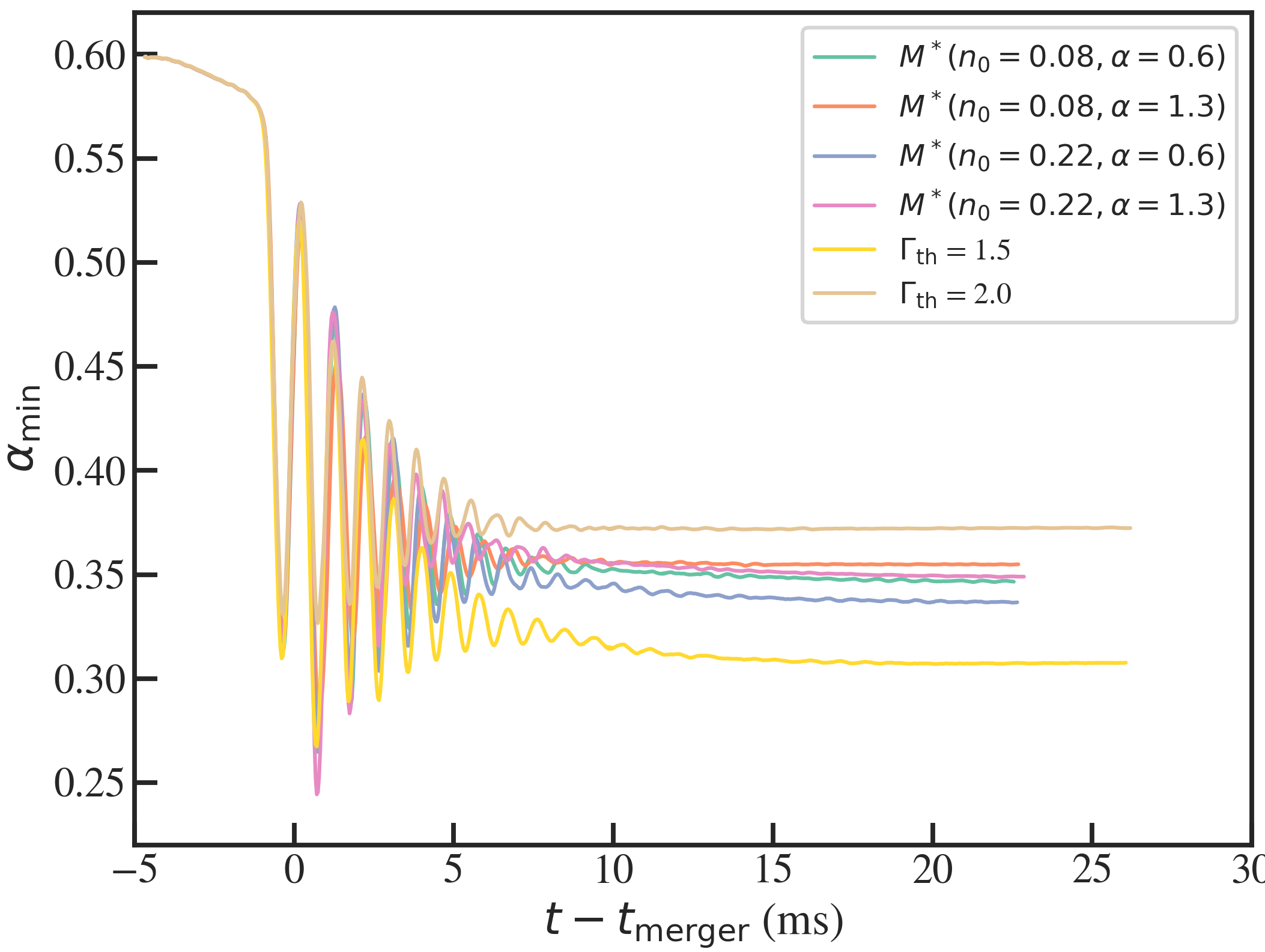}
\caption{\label{fig:lapse}  Minimum lapse, $\alpha_{\rm min}$, as a function of time since merger, for the six thermal treatments considered in this paper. The minimum lapse is approximately
constant at late times, indicating
that the remnant remains stable against collapse until
the end of our simulations.}
\end{figure}

\subsection{Stability and convergence}

 For all the $M^*$-parameter choices and for both constant 
 $\Gamma_{\rm th}$ evolutions, we find that the neutron stars remain
 stable and show no signs of significant 
 heating prior to merger, as is consistent
with previous findings \citep[e.g.,][]{Oechslin2007}. As a result,
all thermal treatments lead to nearly identical inspirals.

The rest mass of the merger remnant is $\sim$3.23$\Ms$, which exceeds
the maximum rest mass for the zero-temperature Kepler sequence of
3.17~$\Ms$. This suggests that the remnant is likely supported by
differential rotation, with the thermal pressure providing additional
support~\citep{Paschalidis2012}, but that the remnant should
eventually collapse. However, we find no signs of collapse by the end
of our evolutions, which last for $\sim$20~ms following the merger for
the $M^*$-EoSs and 25~ms post-merger for the hybrid evolutions.
Figure~\ref{fig:lapse} shows that the minimum lapse function remains
stable at late times, indicating that the remnant has not started
collapsing by the end of these simulations, for all thermal treatments
considered here.

Finally, we also perform evolutions at 1.5625 and 2$\times$ the baseline
resolution for the EoS with $M^*$-parameters
$n_0=0.08$~fm$^{-3}$ and $\alpha$=1.3. We find second-order
convergence of $||\mathcal{H}||$ during the inspiral and for the first
few milliseconds post-merger, which then decays at later times (see
Appendix~\ref{sec:BNSresolution} for more details).

\subsection{Post-merger evolution}
\label{sec:thermal}

\begin{figure*}[ht]
\centering
\includegraphics[width=0.9\textwidth]{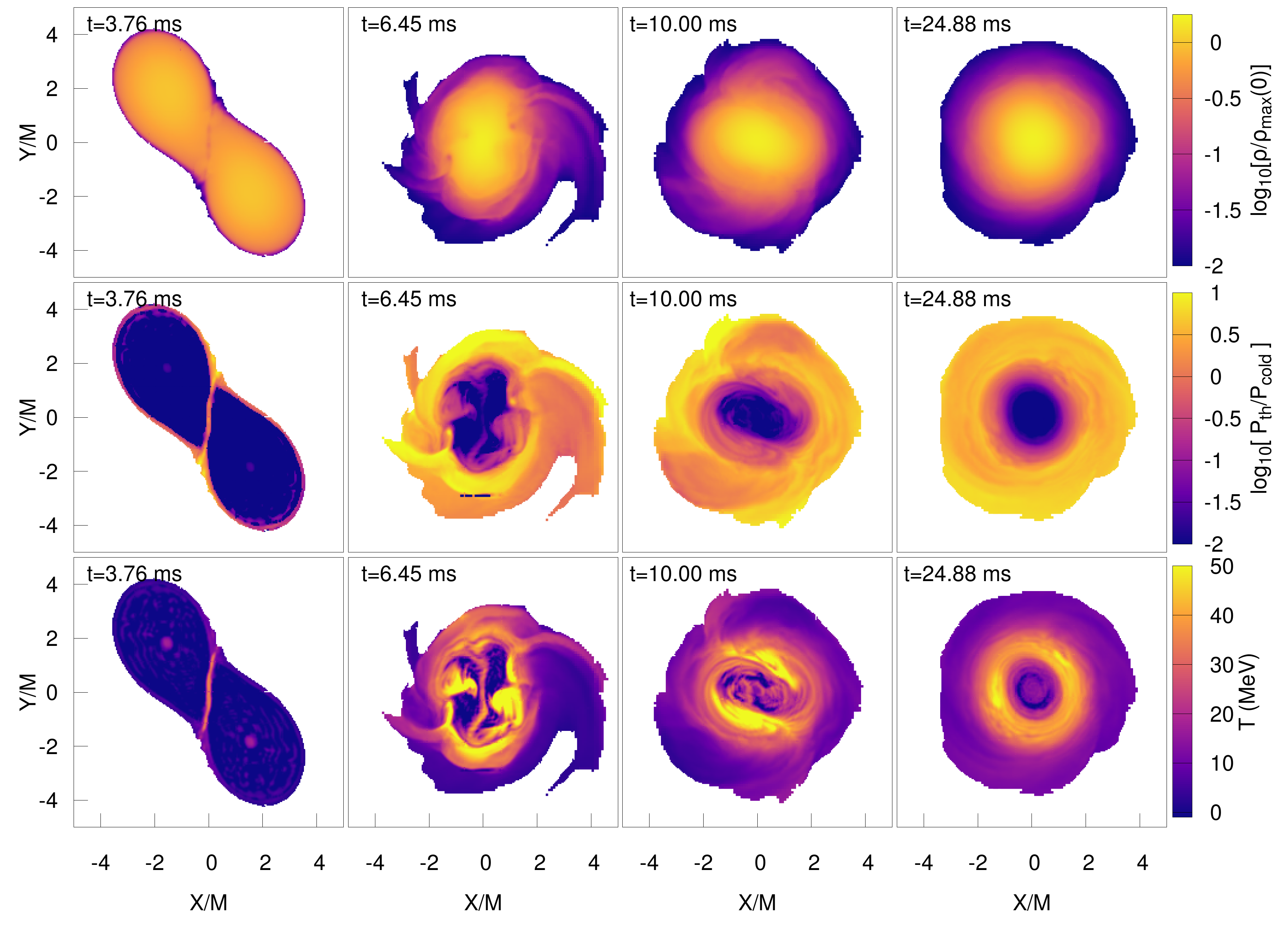}
\caption{\label{fig:profiles_n08a13} Top: Density profile just before and
  at three snapshots after merger, for the $M^*$-EoS with
  $n_0=0.08~\text{fm}^{-3}$ and $\alpha=1.3$. Middle: Thermal pressure
  profile, relative to the cold pressure, at the same times. Bottom:
  Temperature profile, extracted from the density and thermal pressure
  using the microphysical model of Eq.~(\ref{eq:Pth}). All plots only
  include matter with densities above 0.01$\times$ the initial maximum
  rest-mass density, $\rho_{\rm max}(0)$.}
\end{figure*}

\begin{figure*}[ht]
\centering
\includegraphics[width=0.9\textwidth]{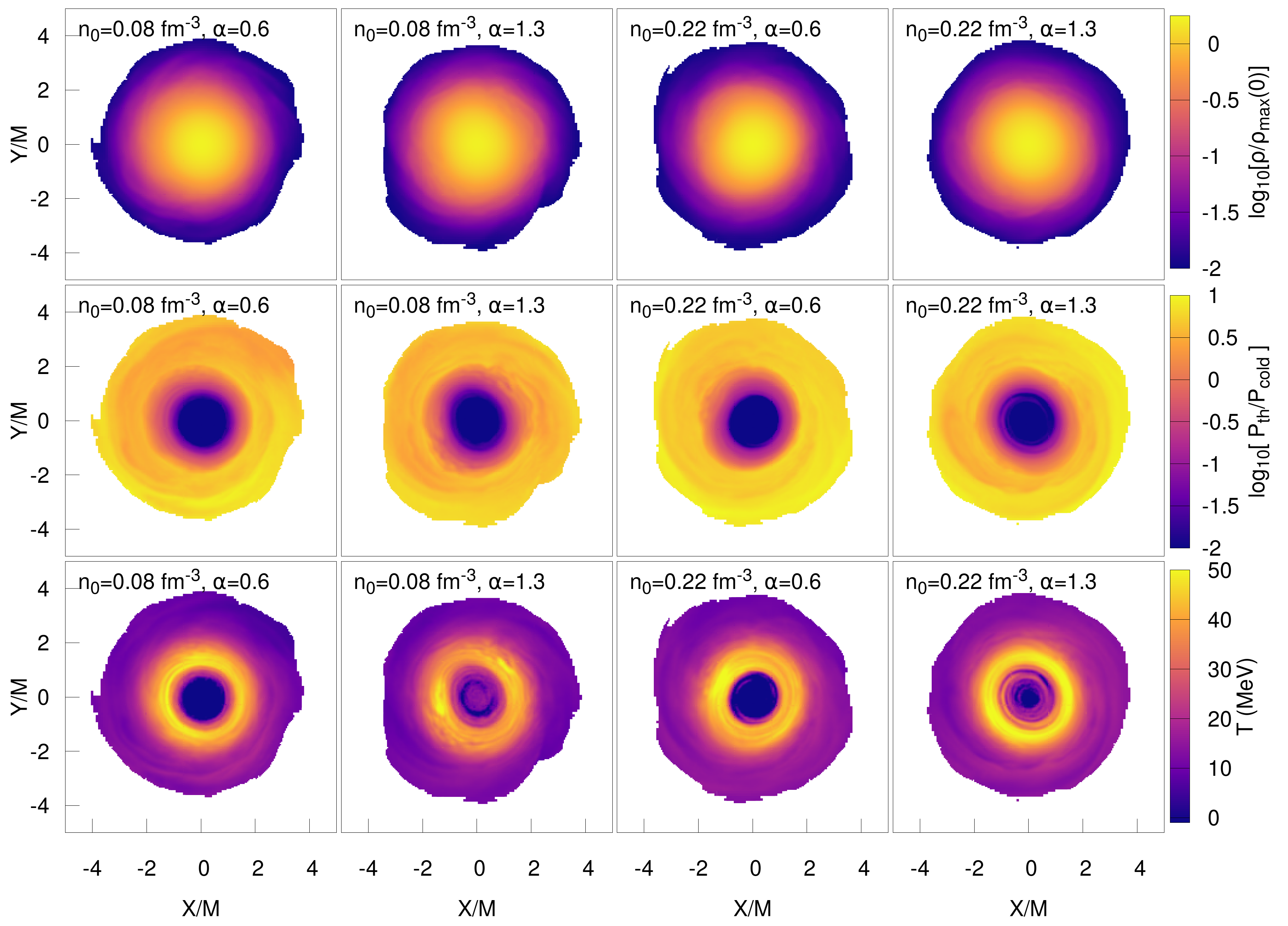}
\caption{\label{fig:profiles_lateT} Late-time ($t=24.88$~ms) profiles
  for each of the $M^*$-EoSs.  Each column corresponds to the specific
  set of $M^*$-EoSs parameters indicated, and the rows show the
  density (top), thermal pressure (middle), and temperature (bottom)
  profiles. We find that the thermal pressure in 
  the outer layers of the star is primarily determined by the value of $n_0$, 
  while the temperature of the inner core ($|X/M|\lesssim 1$) is determined by $\alpha$.}
\end{figure*}

In order to give a qualitative sense of the behavior of the density
and thermal profiles over time, we show 2D snapshots from the
$M^*$-EoS simulation with $n_0=0.08~\text{fm}^{-3}$ and $\alpha=1.3$
in Fig.~\ref{fig:profiles_n08a13}, just prior to merger and at
select times post-merger.  The top row shows 2D rest-mass density
profiles; the middle row shows the thermal pressure relative to the
cold pressure; and the bottom row shows the corresponding temperature,
which is computed from Eq.~(\ref{eq:Pth}). For comparison, we also
show the late-time ($t=24.88$~ms) 2D profiles for all four $M^*$-EoSs
in Fig.~\ref{fig:profiles_lateT}, with different $M^*$-parameters
shown in each column. In each of these figures, we only include matter
with densities above $10^{-2}\times$ the initial central density of
each star, $\rho_{\rm max}(0)$, with lower-density material masked in
white. Additionally, in Figs.~\ref{fig:profiles_n08a13} and
\ref{fig:profiles_lateT}, wherever the thermal pressure is
negative, it is replaced with zero for display purposes (i.e., both zero and
negative thermal pressures are shown as dark purple, to indicate the
matter is ``cold"; see $\S$\ref{sec:Mstar} for further discussion).

From the snapshots shown in Figs.~\ref{fig:profiles_n08a13} and
\ref{fig:profiles_lateT}, several trends emerge. First, we find
evidence of significant heating at supranuclear densities.
Figure~\ref{fig:profiles_n08a13} shows that the stars remain cold
prior to merger, but that the thermal pressure can reach a few tens of
percent of the cold pressure shortly following merger.  At late times,
Fig.~\ref{fig:profiles_lateT} shows that differences persist in the
thermal pressure profile depending on the $M^*$-parameters, with
higher $P_{\rm th}/P_{\rm cold}$ in the outer layers of the remnant
for evolutions with $n_0=0.22$~fm$^{-3}$ than with
$n_0=0.08$~fm$^{-3}$.  However, for all $M^*$ parameters, the very
core of the star (e.g., $|X/M| \lesssim 1$)
remains thermodynamically cold ($P_{\rm th} \lesssim
0.1 P_{\rm cold}$) at late times.

Additionally, in comparing these snapshots, it becomes clear that
small differences in the thermal pressure can translate to large
differences in the temperature profile of the remnant, due to the
$\propto T^2$ dependence in Eq.~(\ref{eq:Pth}), which dominates at
high densities.  From the bottom panel of
Fig.~\ref{fig:profiles_lateT}, we find that larger values of $\alpha$ 
correspond to higher core temperatures at late times.  When
$\alpha$ is large, $M^*$ decays more quickly.
Thus, Fig.~\ref{fig:profiles_lateT} suggests that having a small effective mass
 at the core leads to larger core temperatures.
 This is similar to the findings from
1D CCSN simulations, in which EoSs with a smaller effective mass were
found to produce larger central temperatures in the proto-neutron star
\citep{Schneider2019,Yasin2020}. However, we note that the trend
breaks down at other densities in our merger remnants: that is, it is not generically true
that temperature scales with the local effective mass at every density.

These findings suggest that
the parameters of the $M^*$-approximation play a role
in determining the post-merger thermal profile, with larger $n_0$
contributing to a higher degree of heating in the outer layers, and larger 
$\alpha$ contributing to hotter cores. As a result,
the local neutrino emissivity, and hence the cooling and
ultimate neutrino irradiation of the remnant disk likely will also depend on the
parameters characterizing the finite temperature part of the EoS.

\begin{figure*}[ht]
\centering
\includegraphics[width=\textwidth]{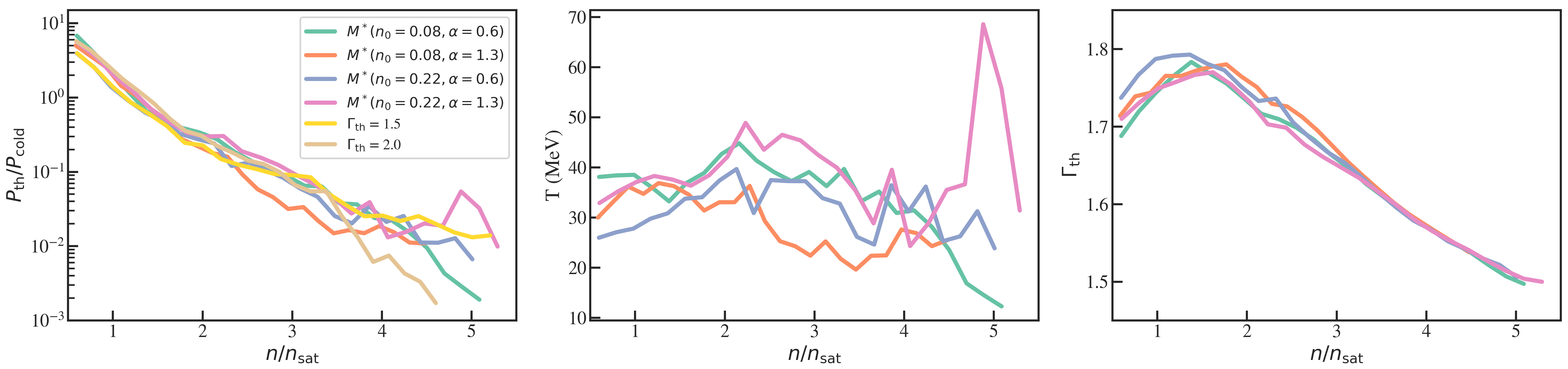}
\caption{\label{fig:PthT}  Characteristic $P_{\rm th}/P_{\rm cold}$ (left), temperature (middle), and thermal index (right) at each density. We define the characteristic quantity as the median of the distribution of values within a particular density bin, at a fixed time just after merger ($t=6.5$~ms). We only extract temperatures for the $M^*$-EoSs, which have a microphysical relationship between $P_{\rm th}$ and $T$.}
\end{figure*}

In order to be more quantitative in our comparison,  we also calculate characteristic
1D profiles of the thermal pressure, temperature, and thermal index just after merger,
when the matter has not yet been redistributed by the differing thermal pressures.
To compute these characteristic quantities, we first bin all grid points along
the equatorial plane at a fixed time ($t=6.5$~ms), using
density bins that are uniformly spaced between 0.5~$n_{\rm sat}$ and the core density.
 Within each density bin, we then compute the distribution of $P_{\rm th}/P_{\rm cold}$,
$T$, and $\Gamma_{\rm th}$, and we take the median value as characteristic.
We show these characteristic values as a
function of the corresponding density bin in Fig.~\ref{fig:PthT}.
 We find that, at the nuclear
saturation density, the thermal pressure can be a few times larger than the cold
pressure, but that it decreases in relative importance at higher densities. At
core densities ($\sim5n_{\rm sat}$),
 the typical thermal pressure is $\lesssim 0.1 P_{\rm cold}$ in
all cases, but the exact value can vary by up to an
order of magnitude at these densities, depending on the thermal treatment.
 The $M^*$ evolution with $n_0$=0.22~fm$^{-3}$ and $\alpha=1.3$
leads to the largest thermal pressure at the core just after merger, whereas the evolution
with  $n_0$=0.08~fm$^{-3}$ and $\alpha=0.6$ produces the coldest core. These thermal pressures 
correspond to core temperatures ranging from nearly 70~MeV to $\sim12$~MeV, respectively.
The other two sets of $M^*$ parameters lead to nearly identical core temperatures, just after merger,
but still differ significantly from each other $M^*$-EoS throughout the rest of the star.

\begin{figure}[ht]
\centering
\includegraphics[width=0.45\textwidth]{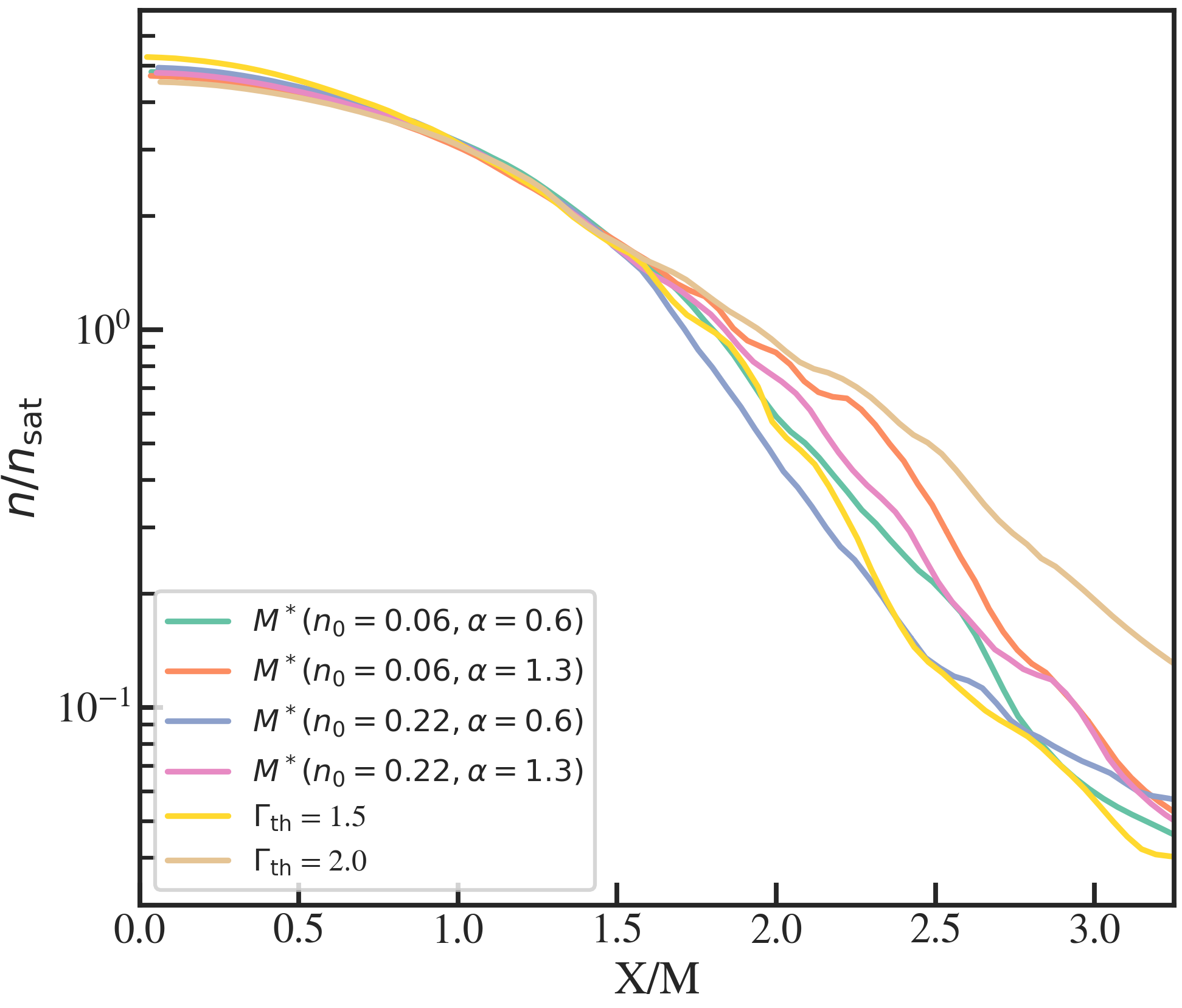}
\caption{\label{fig:dens_profiles} Density profiles along the X-axis
  at late times ($t=24.88$~ms), for the six different thermal
  treatments. Although the remnant starts with a similar density
  profile in all cases, the different thermal pressures in each of the
  six models cause the matter to be redistributed in noticeably
  different ways by late times.}
\end{figure}

The thermal pressure profile just after merger is particularly
interesting to consider, since this governs in part the redistribution
of matter within the remnant and, hence, the post-merger evolution. We
show how the differences in $P_{\rm th}/P_{\rm cold}$ just after
merger influence the resulting remnant structure in
Fig.~\ref{fig:dens_profiles}, where we plot 1D density profiles,
extracted along the X-axis, at the end of our simulations
($t=24.88$~ms). We find small differences in the central density of
the remnant between our various evolutions, with the $M^*$ evolutions
differing by $\lesssim5$\% and the hybrid evolutions differing by
$\sim15$\% from one another.  The late-time radial extent of the star
differs more significantly depending on the thermal treatment, with 
large values of $\alpha$ or large $\Gamma_{\rm th}$
leading to a more extended mass distribution. 
Although coordinate size is not a gauge-invariant quantity,
Fig.~\ref{fig:dens_profiles} is suggestive that differences in the
thermal treatment may influence the final compactness of the remnant.

\begin{figure}[ht]
\centering
\includegraphics[width=0.45\textwidth]{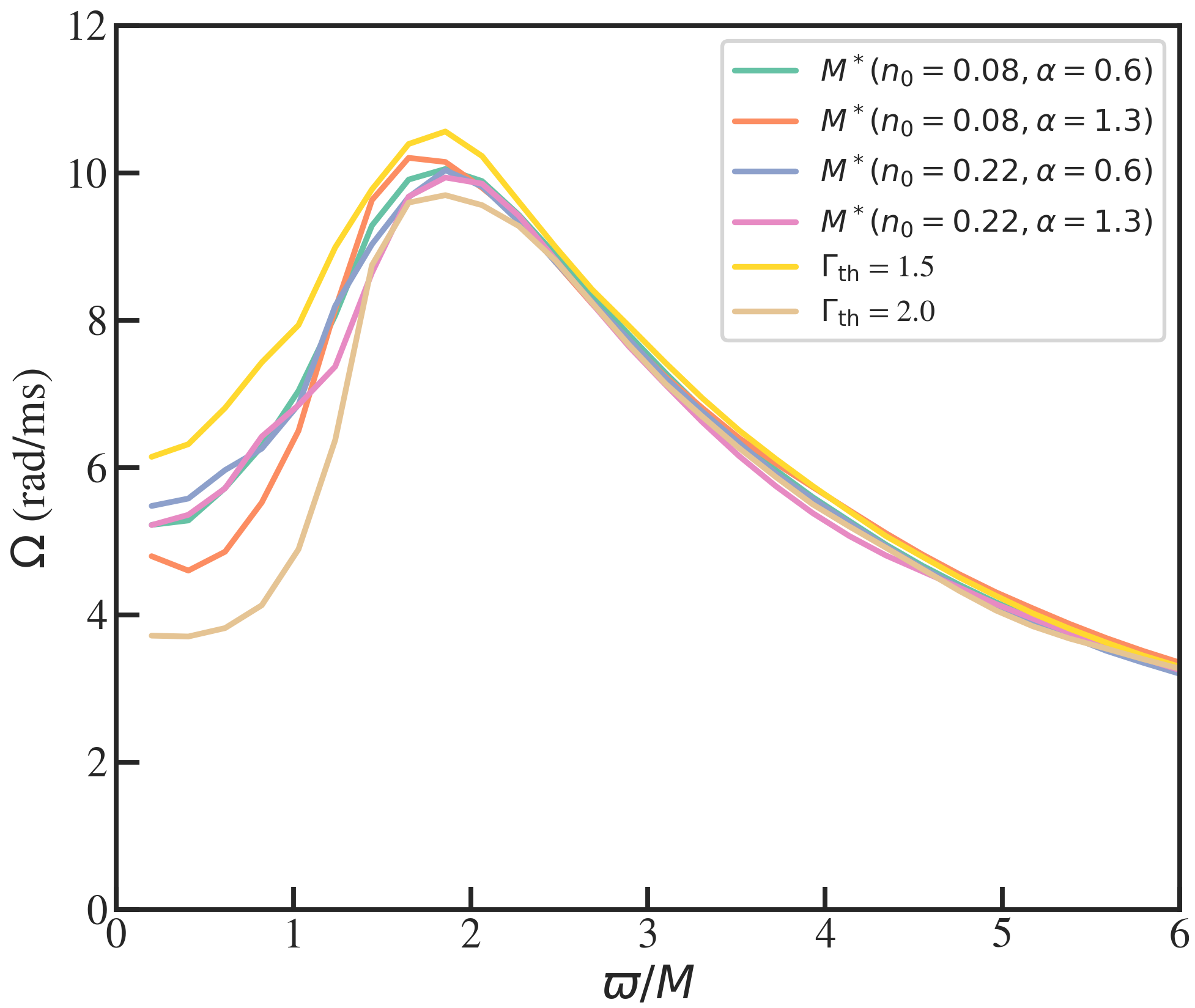}
\caption{\label{fig:angVel} Azimuthally-averaged angular velocity
 profiles as a function of the cylindrical coordinate radius, on the equator
 of the remnant at late times ($t=24.88$~ms).  The different thermal treatments lead
 to up to 60\% differences in the core angular velocities and 10\% differences in the
 peak angular velocities, with a reduced range found between the $M^*$-parameters.}
\end{figure}

Finally, Fig.~\ref{fig:angVel} shows the azimuthally-averaged angular
velocity, $\Omega = v^{\phi}$, as a function of the cylindrical
coordinate radius, $\varpi = \sqrt{X^2+Y^2}$, for each of the thermal
treatments. These profiles are calculated on the equator of the
remnant at the end of the evolution ($t=24.88$~ms). We find that the
angular velocity profile is sensitive to the finite-temperature part
of the EoS, with core angular velocities that differ by up to 60\% and
peak angular velocities that differ by up to 10\% for the six thermal
treatments explored here.  Among only the $M^*$-EoSs, the range of
angular velocities is smaller, with differences of up to $\sim$14\%
and 3\% in the the core and peak angular velocities,
respectively. Trends with particular $M^*$-parameters are harder
to identify in these velocity profiles, but we note that the evolution 
with $n_0=0.08$~fm$^{-3}$ and $\alpha=1.3$ leads to the lowest
core velocity and the largest peak velocity. The other $M^*$-parameter
choices lead to more similar velocity profiles.
In all cases, the overall shape of the angular velocity
profile remains the same as has been found in earlier studies
(see~\cite{2019PhRvD.100l4042E} and discussion therein).

\subsection{Gravitational wave signal}
\label{sec:gw}
\begin{figure*}[!ht]
\centering
\includegraphics[width=\textwidth]{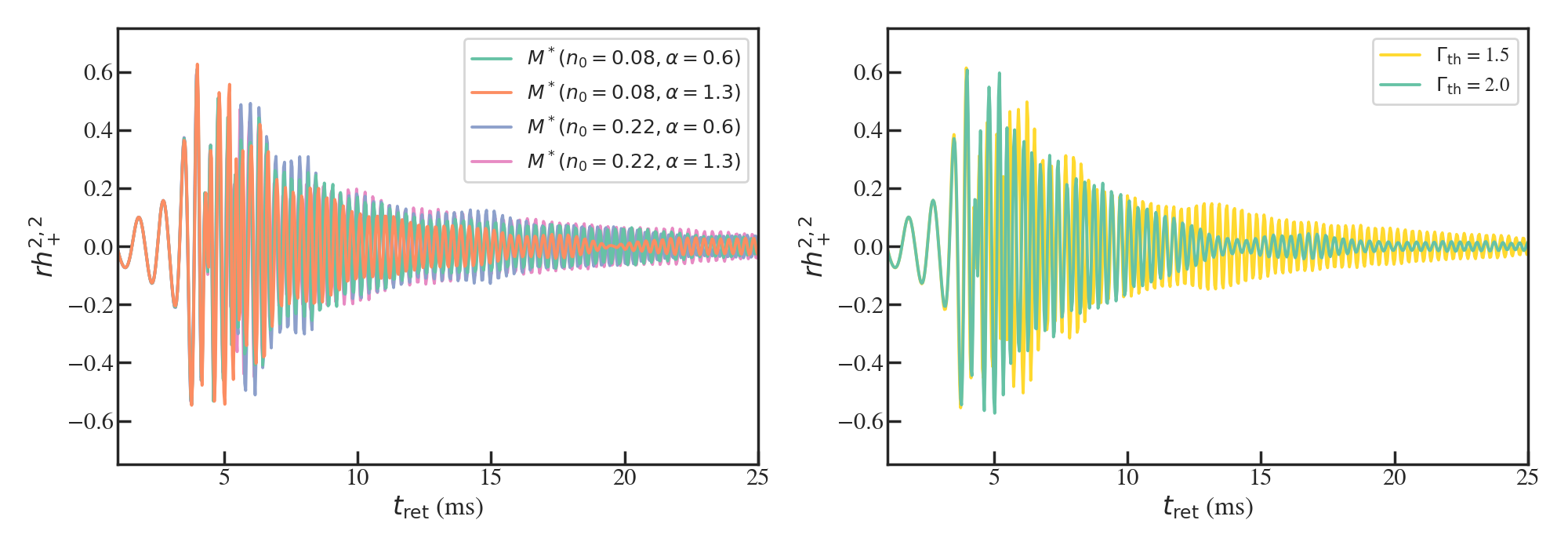}
\caption{\label{fig:strain}
  Gravitational wave strain for the $\ell=m=2$ mode, for the six
  different evolutions, as a function of the retarded time.
  The gravitational wave signals from the
  $M^*$-EoS evolutions are shown in the top panel; the strains from
  the hybrid evolutions are shown in the bottom panel. In all cases,
  the inspiral phase is nearly identical, but we find significant
  differences in the post-merger gravitational waves.}
\end{figure*}

We extract the GW signal, as discussed in $\S~$\ref{sec:diag}, for
each evolution and show the resulting strains in
Fig.~\ref{fig:strain}, for the $\ell=m=2$ mode. We separate the four
$M^*$-EoS evolutions (left) from the two constant-$\Gamma_{\rm th}$
evolutions (right) for visual clarity. In all cases, the inspiral
waveform is nearly identical, with a characteristic time to merger of
4.7~ms.  By contrast, we find significant differences in the
post-merger gravitational waves across all six thermal
treatments. Figure~\ref{fig:strain} shows differences not only between
the amplitudes of the post-merger strains, but also between the beat
frequencies of the decaying signals, suggesting that the post-merger
oscillation frequencies also depend on the thermal treatment.
 
Many previous studies have found evidence of
empirical correlations between
 the oscillation frequencies of the the post-merger GW 
 signal and the neutron star radius or stellar compactness 
 \citep[for reviews, see][]{Baiotti2017,Paschalidis2017,Bauswein2019}. 
 These correlations make it theoretically possible to constrain the properties 
 of the initial, cold neutrons through the measurement of the
 post-merger GW power spectrum. Using these types of relationships, it has 
 been estimated that Advanced LIGO may constrain the radius to within
  0.429~km for a nearby ($\lesssim30$~Mpc) event \citep{Clark2016}. 
  It may be possible to get even smaller errors by coherently stacking 
  post-merger spectra from multiple events with third-generation facilities, 
  at which point systematic errors in the universal relations may dominate 
  the error budget \citep{Yang2018}. However, these estimates do not
  explicitly account for the uncertainties in the finite-temperature part
 of the EoS, which are hinted at in Fig.~\ref{fig:strain} and which
 may be important to take into account in order to extract
 precision radius estimates from such spectral features.
 
 On the other hand, if the cold EoS can be pinned down from other
 observations -- e.g., from the NSNS inspiral or from X-ray
 observations -- then any remaining dependence of the the post-merger
 oscillation frequencies on the $M^*$-parameters could potentially be
 used as an exciting new probe of the finite-temperature part of the
 nuclear EoS.  We leave a more detailed exploration of the dependence
 of the post-merger GWs on the various $M^*$-parameters to future
 work.

\subsection{Ejected mass}

\begin{figure}[ht]
\centering
\includegraphics[width=0.45\textwidth]{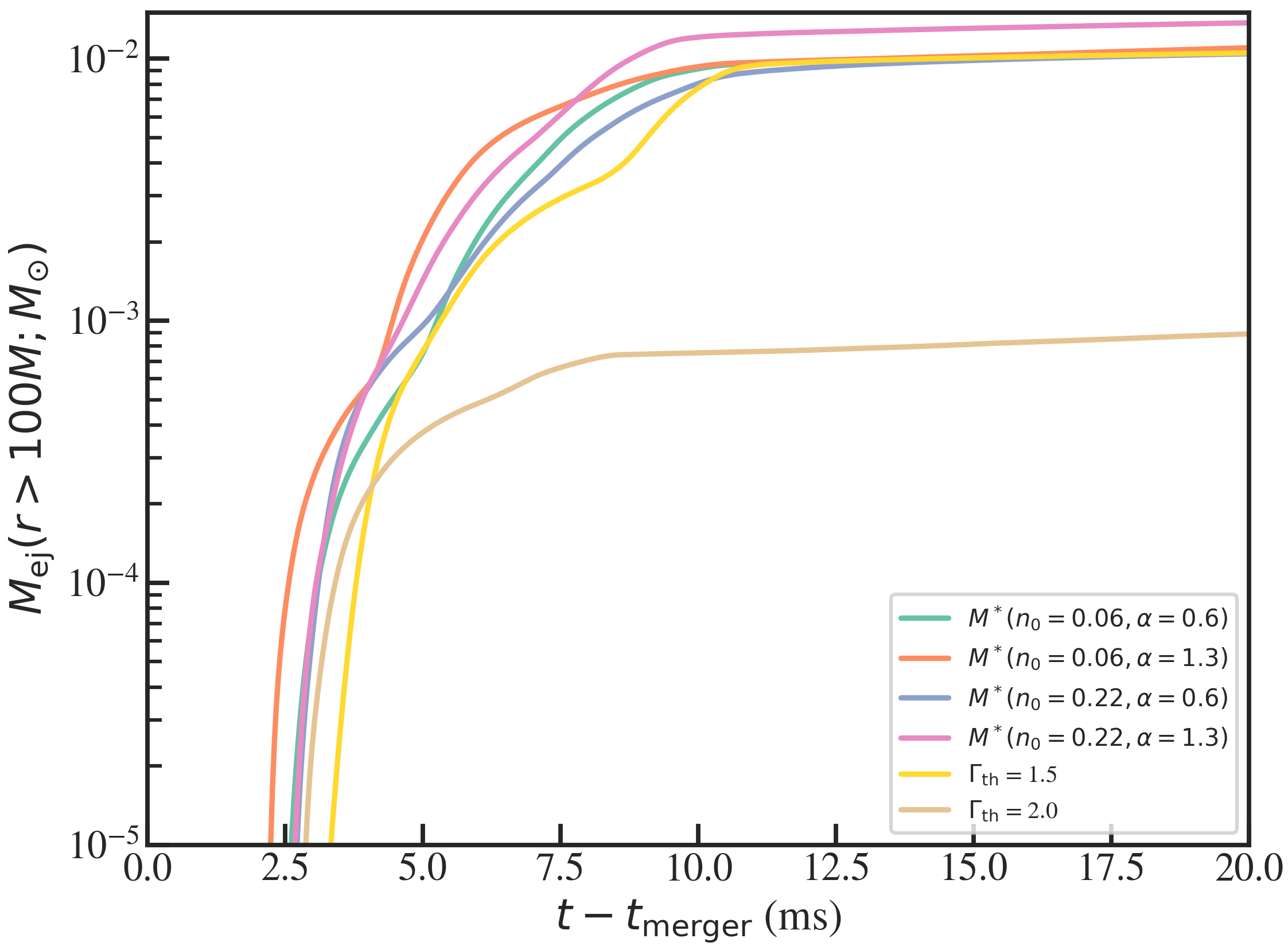}
\caption{\label{fig:Mej}  Ejected matter over time for the six EoSs considered in this work. The rapid rise is generated by the fast-moving ejecta, while the increase in $M_{\rm ej}$ at late times is caused by the slow-moving tail of the distribution of ejected matter.}
\end{figure}

Finally, we also calculate the ejected mass for each of our evolutions
via Eq.~(\ref{eq:Mej}) for a sphere with radius $100 M$.
Figure~\ref{fig:Mej} shows the ejecta over time. We find a rapid rise
in $M_{\rm ej}$ for the first $\sim$10~ms post-merger. For the
$M^*$-EoSs in particular, we find that the fastest 10$^{-4}~\Ms$ of
ejecta have speeds of up to $\sim 0.5$~$c$ for the $M^*$ evolutions,
while the fastest ejecta in the hybrid evolutions tend to be 
somewhat slower, with speeds of up to $\sim 0.4$~$c$.

We extract the amount of ejecta rest mass at the end of our
simulations (20~ms post-merger). While $M_{\rm ej}$ is still slowly
increasing at late times, due to the slow-moving tail of the
distribution of ejecta, we use this value to facilitate an approximate
comparison of $M_{\rm ej}$ between the different thermal treatments.
We also note that the integration to compute $M_{\rm ej}$ introduces
some error, which we estimate by comparing the extracted values of
$M_{\rm ej}$ between a low- and high-resolution evolution. Based on
this comparison, we estimate that the error in our reported values of
$M_{\rm ej}$ is $\sim$150\% (see Appendix~\ref{sec:BNSresolution} for
further details). Higher resolution is necessary for more accurate
estimates.

As shown in Fig.~\ref{fig:Mej}, we find that $M_{\rm ej}$ varies from
$\sim1.0-1.3\times10^{-2}~\Ms$ for the $M^*$-evolutions, which implies
a range much smaller than our estimated error.  Thus, although we do
find some dependence of $M_{\rm ej}$ on the $M^*$-parameters, 
the differences are not numerically significant,
at least for the particular cold EoS, binary parameters, and
resolutions explored here. In contrast, there is a factor-of-8
difference between $M_{\rm ej}$ for the $\Gamma_{\rm th}=1.5$ and
$\Gamma_{\rm th}=2$ evolutions, with the latter producing
significantly less ejecta.

In \cite{Hotokezaka2013} it was previously found that larger
$\Gamma_{\rm th}$ can lead to suppressed ejecta. In that work, the
authors suggested that $M_{\rm ej}$ depends on $\Gamma_{\rm th}$ in two
competing ways. On the one hand, a larger value of $\Gamma_{\rm th}$
leads to more efficient shock heating, which acts to increase the
amount of matter ejected.  However, the remnant is also less compact
for large $\Gamma_{\rm th}$ 
 and, accordingly, has a slower rotational velocity
(as shown in our Figs.~\ref{fig:dens_profiles} and \ref{fig:angVel}).
This reduces the torque that the remnant exerts onto the
surrounding material and, as a result, less matter becomes unbound
\citep{Hotokezaka2013}. Thus, somewhat counterintuitively, a large
$\Gamma_{\rm th}$ can indeed lead to suppressed ejecta. We leave further
analysis of the properties of the ejecta to future work.

\section{Conclusions}
In this paper, we have implemented a new prescription for studying
finite-temperature effects in binary neutron star mergers, using a
two-parameter approximation of the particle effective mass. This is
the first implementation of parametric finite-temperature effects that
include the effects of degeneracy in a compact binary merger
simulation. We tested this new prescription in rotating, single stars
that are initially cold or that initially have a non-zero temperature
gradient, as well as in several binary evolutions, and we find that
the EoS can support stable stars over long timescales.

We also performed a parameter study to explore a broad range
of $M^*$ values in a series of NSNS merger simulations. We considered
four sets of $M^*$-parameters, as well as two constant-$\Gamma_{\rm
  th}$ values in order to provide a basis of comparison for the new
$M^*$ results. While the inspiral portion of the merger is virtually
identical for all six thermal treatments, we find significant
differences in the post-merger evolution for the different thermal
prescriptions. Depending on the thermal treatment, we find up to an
order of magnitude difference in the characteristic $P_{\rm th}/P_{\rm
  cold}$ at core densities just after the merger. As a result of these
differences in the post-merger thermal profiles, the mass distribution
of the remnant can also vary significantly by the end of our
simulations. 

Perhaps most
interestingly, from an observational point of view, are the
differences that can emerge in the post-merger GW signal. We find
that the post-merger GW strain is sensitive to the particular choice 
of $M^*$-parameters. We plan to further study this dependence 
in future work. If the post-merger GW spectrum does indeed
depend on the parameters of $M^*$, 
as our findings hint at, then observations of post-merger GWs 
may one day offer a new window into the properties of dense matter at
non-zero temperatures. 

Finally, it is worth noting that the relative importance of thermal 
effects may change for binaries with different total mass, mass ratio,
underlying cold EoS (and, hence, stellar compactness), and
potentially also with the added presence of magnetic fields.
For example, we expect that the dependence
of merger properties on the $M^*$-parameters
will become stronger for softer EoSs, which predict more compact stars.
More compact stars are expected to collide at higher velocities,
leading to stronger shock heating and an enhanced thermal pressure. 
Combined with the lower cold pressure of the softer EoS, 
we expect the thermal pressure may play a more important role
in such mergers. We leave the study of such effects to future work.  

\begin{acknowledgments} We are indebted to Antonios Tsokaros for
permission to use initial data he generated with the {\tt COCAL} code
for other projects. We are grateful to William East and Frans
Pretorius for providing us with their generic primitives recovery
routine. We would like to thank Dimitrios Psaltis, Gabriele Bozzola,
Erik Wessel, Ryan Westernacher-Schneider, and Elias Most for useful
conversations related to this work. CR was partially supported during
this project by NSF Graduate Research Fellowship Grant DGE-1746060, as
well as by a joint postdoctoral fellowship at the Princeton Center for
Theoretical Science, the Princeton Gravity Initiative, and as a John
N. Bahcall Fellow at the Institute for Advanced Study.  This research
was in part supported by NSF Grant PHY-1912619 to the University of
Arizona and by NSF PIRE grant 1743747.  The simulations presented in
this work were carried out in part on the {\tt Ocelote} and {\tt
  ElGato} clusters at the University of Arizona, as well as on the
{\tt Stampede2} cluster at the Texas Advanced Computing Center, under
XSEDE allocation PHY190020.

\end{acknowledgments}

 \appendix

\section{Single star test results}
\label{sec:singlestars}

In this appendix, we describe the key results from our single star
evolutions in dynamical spacetimes. For both zero-temperature single
stars and single stars initialized with $P_{\rm th}/P_{\rm
  cold}=0.1$, we perform evolutions for $\sim5~t_{\rm dyn}$, where
$t_{\rm dyn}=1/\sqrt{\rho_{b,c}}$ is the dynamical timescale and
$\rho_{b,c}$ is the central rest mass density.  Each set of initial
data is evolved with the hybrid approximation with $\Gamma_{\rm
  th}=1.66$, as well as with the $M^*$-thermal treatment with
$n_0=0.12$~fm$^{-3}$ and $\alpha=0.8$.

While the hybrid EoS has been well tested within the Illinois
spacetime + GRMHD code in previous studies
\citep[e.g.,][]{Etienne2010,Etienne2015}, we include the test results
here again, in order to validate our implementation of the primitive
recovery scheme of \cite{East2012} into our code, as well as to
provide a basis of comparison for the $M^*$-EoS results.

\begin{figure}[ht]
\centering
\includegraphics[width=0.45\textwidth]{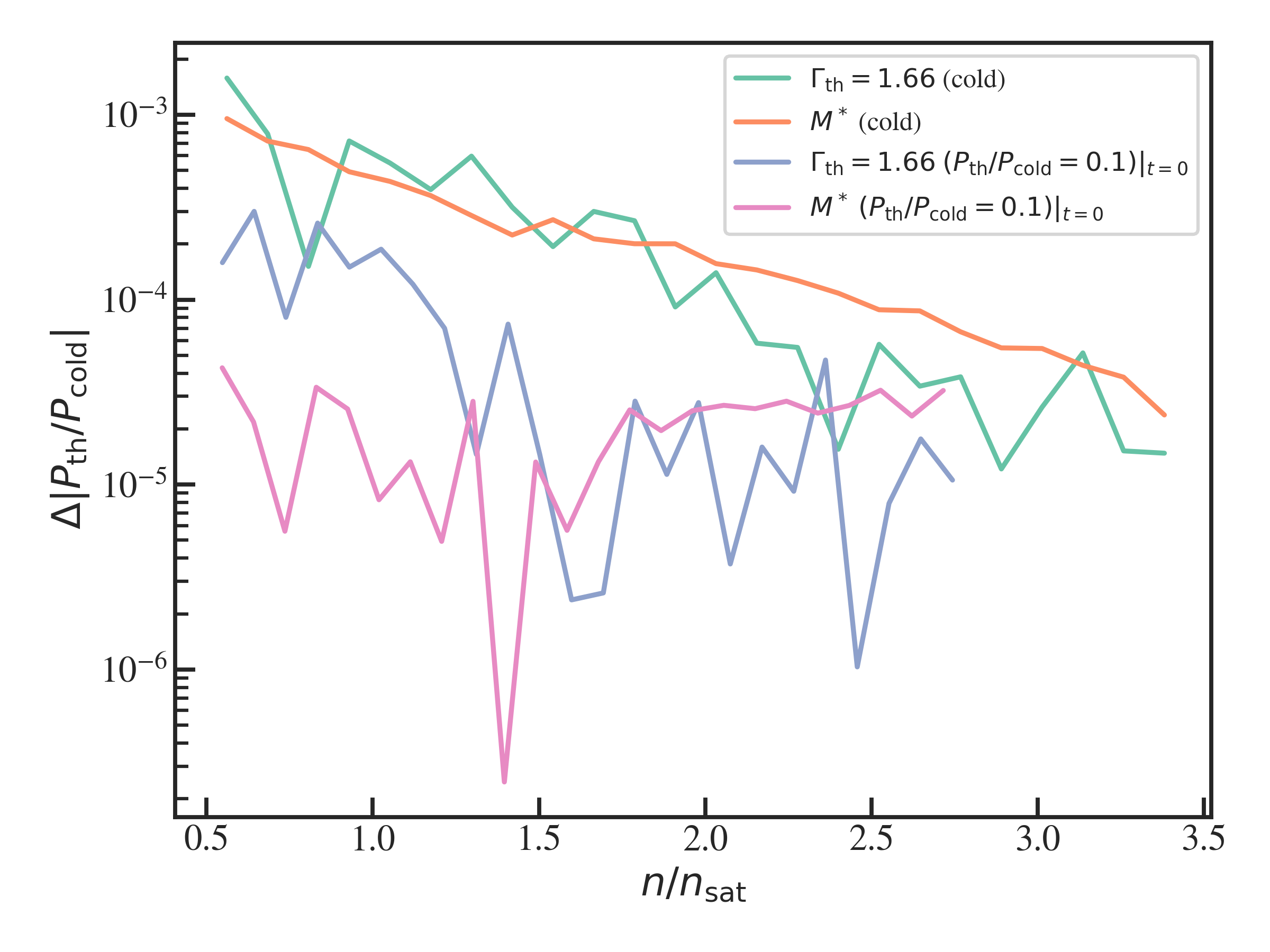}
\caption{\label{fig:avgPthSingles} Total change in the characteristic
  thermal pressure at each density, between the initial and final time
  steps (i.e., $|P_{\rm th}/P_{\rm cold}(t=0) - P_{\rm th}/P_{\rm
    cold}(5t_{\rm dyn})|$). All profiles correspond to the
  highest-resolution evolutions. Both the hybrid and $M^*$ evolutions
  maintain the initial thermal profile to within $\lesssim$1 part in
  10$^3$ at supranuclear densities, for both zero- and
  finite-temperature initial data. }
\end{figure}

For both single star tests, we find that the $M^*$-EoS is able to maintain the initial thermal profile, with no evidence of spurious heating. Figure~\ref{fig:avgPthSingles} shows the change in the characteristic value of $P_{\rm th}/P_{\rm cold}$ from the beginning to the end of the simulation, for each of the single star tests considered. We calculate the characteristic value of $P_{\rm th}/P_{\rm cold}$ in each density bin at each time, as in $\S$\ref{sec:thermal}. Over the $\sim5$ dynamical timescales that were evolved, $P_{\rm th}/P_{\rm cold}$ changes by $\lesssim10^{-3}$ at supranuclear densities.  Additionally, we find that the $M^*$-EoS performs comparably well to the hybrid approximation at maintaining either a zero-temperature or fixed thermal profile.

\begin{figure*}[ht]
\centering
\includegraphics[width=0.9\textwidth]{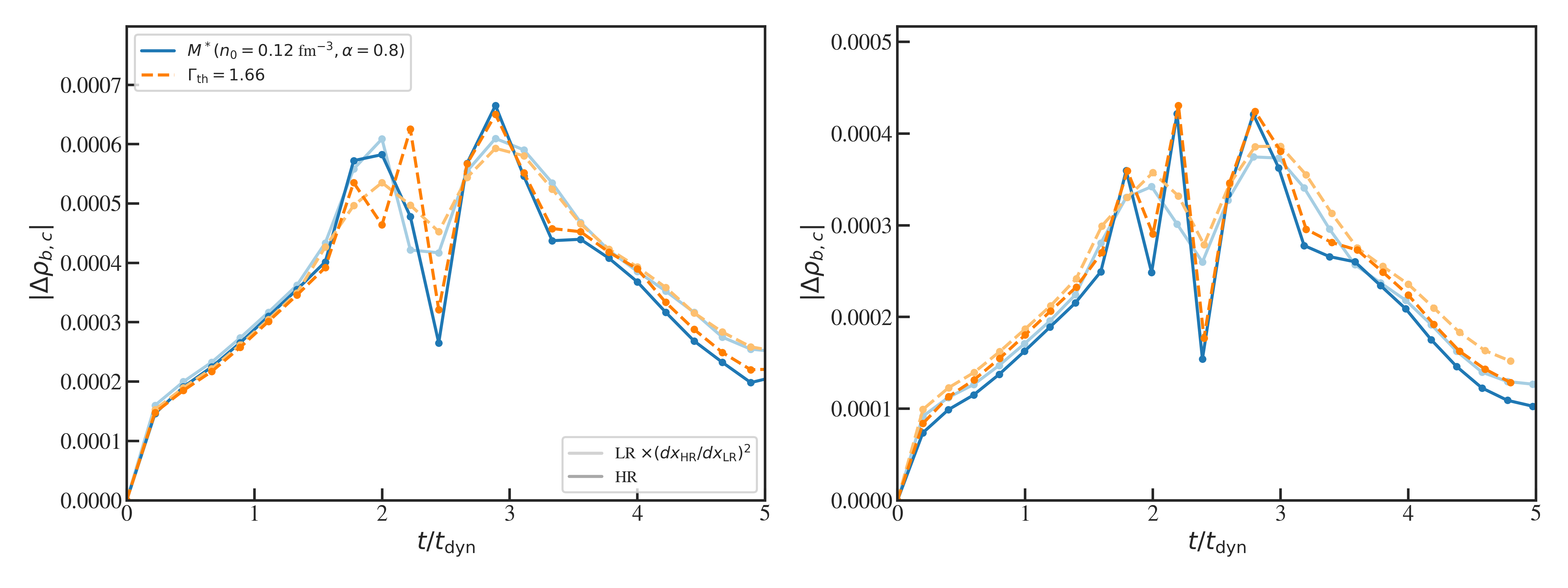}
\caption{\label{fig:T0_diag}  Left: Change in the central rest mass density over five dynamical timescales, for the cold rotating star tests. The blue line represents the tests evolved with the $M^*$-EoS, while the orange dashed line represents the hybrid evolution. The lighter shade corresponds to the low-resolution evolution (which has been scaled to show 2nd-order convergence), while the darker shade indicates the high-resolution evolution. Right: Same as the left panel, but for the rotating star tests with a non-zero initial temperature profile set by $P_{\rm th}/P_{\rm cold}=0.1$. The convergence behavior is identical to the cold evolution. }
\end{figure*}

\begin{figure*}[ht]
\centering
\includegraphics[width=\textwidth]{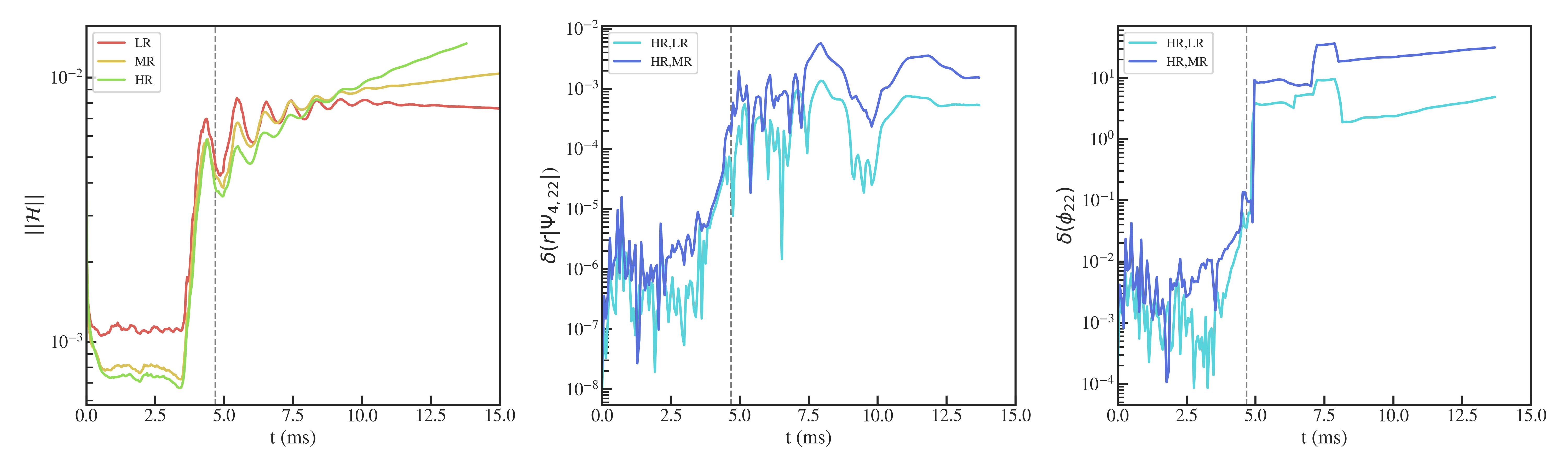}
\caption{\label{fig:Haml} Left: L2 norm of the Hamiltonian constraint
  violation, at three different resolutions, for the evolution with
  $M^*$-parameters $n_0=0.08~\text{fm}^{-3}$ and $\alpha=1.3$. Time of
  merger for the high-resolution evolution is marked with the vertical
  dashed line ($t_{\rm merger}=4.7$~ms). There is second-order
  convergence during the inspiral and for a short period post-merger,
  but that convergence decays at later times. Middle: Self-convergence
  of the amplitude of $\psi_4$ for the $\ell=m=2$ mode, for the same
  evolution. The results have been scaled to show second-order
  convergence. Right: Self-convergence of the phase of $\psi_4$ for
  the $\ell=m=2$ mode, for the same evolution, again scaled to show
  second-order convergence. As with the Hamiltonian constraint, we
  find second-order convergence at early times, which disappears after
  the merger. }
\end{figure*}

In order to monitor the stability of the stars, we track the
time-evolution of the quantity, $\Delta\rho_{b,c}$, which represents
the change in central rest mass density relative to the value at
$t=0$, and which is expected to converge to zero at second-order with
increasing resolution. We show this quantity in Fig.~\ref{fig:T0_diag}
for the zero-temperature evolutions (left panel) and constant $P_{\rm
  th}/P_{\rm cold}$ thermal profile (right panel). In both cases, the
low-resolution quantities have been scaled to show 2nd-order
convergence.

We find the anticipated 2nd-order convergence in
$\Delta\rho_{b,c}$ over time, in both the cold and finite-temperature
evolutions. The results are virtually indistinguishable between the
$M^*$ and hybrid evolutions, thus providing additional validation of
the numerical implementation of the $M^*$-EoS.

\section{Resolution study for binary evolutions}
\label{sec:BNSresolution}

We now present the results of the resolution study for the NSNS merger
simulations, evolved with the $M^*$-EoS with $n_0=0.08$~fm$^{-3}$ and
$\alpha=1.3$. The baseline (low) resolution is described in
$\S$~\ref{sec:grid}; the medium resolution is 1.5625$\times$ finer;
and the high resolution is $2\times$ the baseline resolution. These
resolutions correspond to $\sim$100, 156, and 200 grid points across
the diameter of each initial star, respectively.

Figure~\ref{fig:Haml} shows the convergence of the L2 norm of the
Hamiltonian constraint in the left panel, as well as the
self-convergence of the amplitude and the phase of $\psi_4$ in the
middle and right panels, respectively. The Hamiltonian
constraint violation is expected to converge to zero, with increasing
resolution. For the amplitude and the phase of $\psi_4$, we determine if 
there is self-convergence by comparing to the highest- resolution evolution.
Self-convergence at second-order requires
\begin{equation}
\frac{Q_{\rm LR} - Q_{\rm HR}}{Q_{\rm MR}-Q_{\rm HR}} = \frac{ \left( \Delta x_{\rm LR}/\Delta x_{\rm HR} \right)^2-1}{\left( \Delta x_{\rm MR}/\Delta x_{\rm HR} \right)^2-1},
\end{equation}
where $Q$ is the quantity of interest; LR, MR, and HR indicate low, medium,
 and high resolutions, respectively; and $\Delta x$ is the grid spacing of each resolution. 
 Rearranging this expression, second-order convergence equivalently implies
\begin{multline}
\label{eq:selfConv}
\left( Q_{\rm LR} - Q_{\rm HR} \right) \left[ \left( \Delta x_{\rm MR}/ \Delta x_{\rm HR} \right)^2-1 \right] = \\
\left( Q_{\rm MR}-Q_{\rm HR} \right) \left[ \left(  \Delta x_{\rm LR}/ \Delta x_{\rm HR} \right)^2-1 \right].
\end{multline}
These scaled, differential quantities are plotted in the middle and right panels
of Fig.~\ref{fig:Haml}, for the amplitude and phase of $\Psi_4$. The
 left-hand side of Eq.~\ref{eq:selfConv} is plotted in light blue in Fig.~\ref{fig:Haml},
while the right-hand side is plotted in dark blue. 
The degree to which these two sets of lines agree
indicates how close the results are to exhibiting second-order convergence.

We find second-order
convergence-to-zero in $||\mathcal{H}|| $ during the inspiral and for
the first few milliseconds post-merger. At later times, the
convergence order decays significantly. We likewise find second-order
convergence in both the amplitude and the phase of $\psi_4$ during the
inspiral, which also decays at late times.

In order to understand the loss of convergence at late times, we
performed an additional medium-resolution evolution for the
$\Gamma_{\rm th}=1.5$ EoS. In comparing $||\mathcal{H}||$ for the low-
and medium-resolution evolutions with the hybrid EoS, we find
qualitatively similar behavior to what is shown in the left panel of
Fig.~\ref{fig:Haml} -- with second-order convergence at early times
which then disappears within a few milliseconds post-merger. Although
the turbulent nature of the post-merger evolution makes it very
difficult to achieve convergence post-merger, we suspect that the
decay of convergence at late times found for both the hybrid and $M^*$
thermal treatments stems from discontinuities in the piecewise
polytropic representation of the cold EoS. This is further supported
from the fact that our cold $\Gamma=2$ isolated stellar evolutions
exhibit approximate second order convergence as expected.  NSNS merger
simulations performed with different codes have also found a lack of
convergence in the post-merger phase when modeling the cold EoS with
piecewise polytropes \citep[e.g.,][]{Paschalidis2015,East2016a},
lending support to the hypothesis that the issue may stem from the
piecewise polytropes.  We plan to investigate this issue further in
future work.
  
\begin{figure}[ht]
\centering
\includegraphics[width=0.45\textwidth]{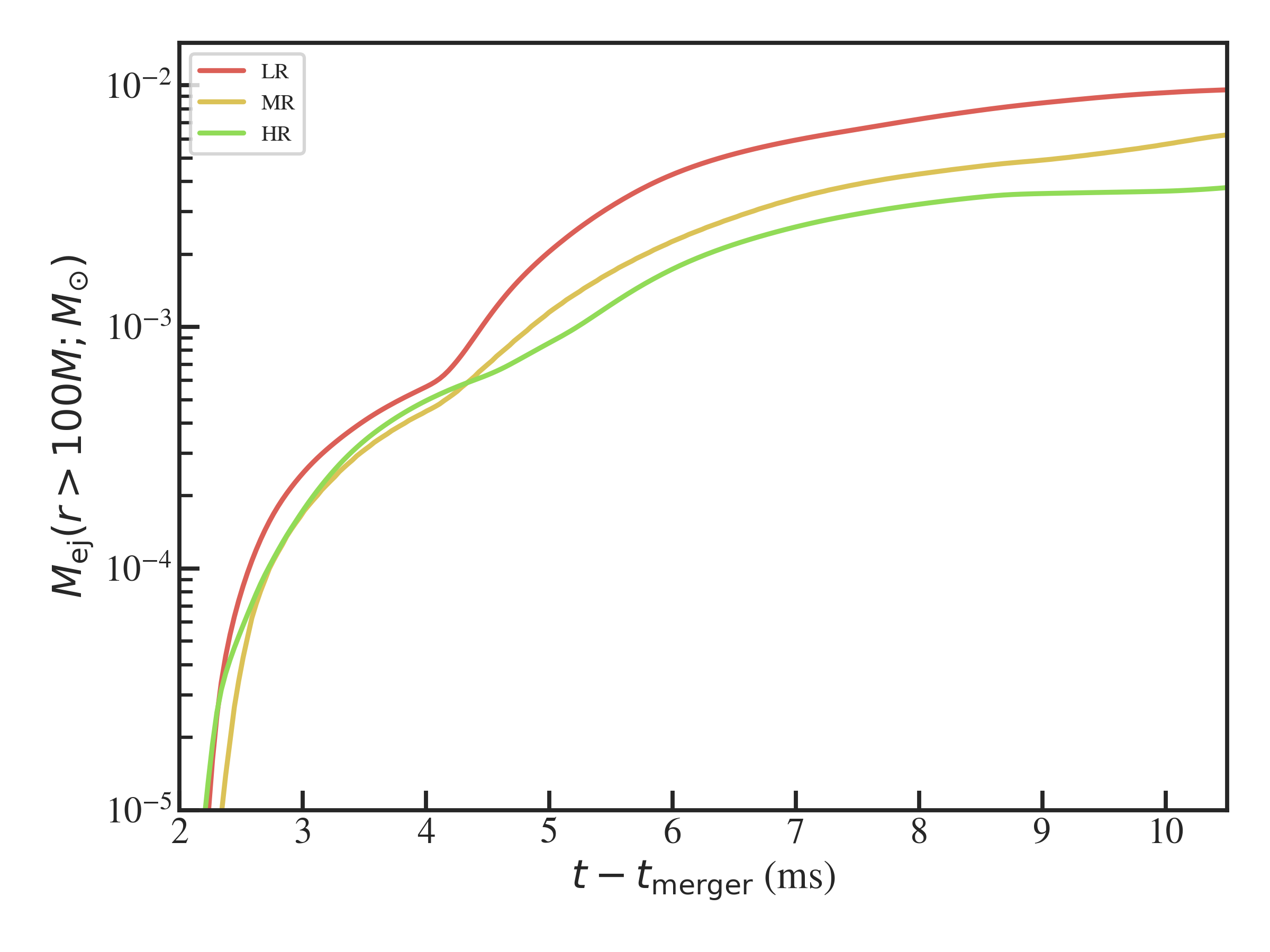}
\caption{\label{fig:MejRes} Ejecta mass for the low-, medium-, and high-resolution evolutions for the $M^*$-EoS with $n_0=0.08$~fm$^{-3}$ and $\alpha=1.3$.  At 10~ms post-merger, the fractional difference in $M_{\rm ej}$ is $\sim150$\% between the lowest and highest resolutions.}
\end{figure}

Finally, we compare the ejecta mass for the low-and high-resolution
evolutions in Fig.~\ref{fig:MejRes}. The differences in the
characteristic speed of the fastest ejecta are negligible between the
three resolutions studied here. However, the total value of $M_{\rm
  ej}$ differs more significantly between the resolutions.  Because
the higher-resolution cases are only evolved for $\sim$10~ms
post-merger, we are unable to extract a late-time value for $M_{\rm
  ej}$ as we did in $\S$~\ref{sec:results}. However, $M_{\rm ej}$ is
starting to asymptote at 10~ms post-merger for all three resolutions,
thereby allowing us to make a reasonable comparison. The values
of $M_{\rm ej}$ extracted in this way indicate 1.5-order convergence.
We note that, although the overall convergence 
of the code decays at late times after the merger, the ejecta are 
launched at the merger and, hence, still exhibit convergence
and can be used to make a reasonable error estimate.
We find a fractional error between the low- and high- resolution values
of $\sim$150\%. 
 
\bibliography{gwthermal}
\bibliographystyle{apsrev4-1}

\end{document}